\documentclass[fleqn,usenatbib]{mnras}

\usepackage[T1]{fontenc}

\DeclareRobustCommand{\VAN}[3]{#2}
\let\VANthebibliography\thebibliography
\def\thebibliography{\DeclareRobustCommand{\VAN}[3]{##3}\VANthebibliography}

\usepackage{graphicx}	
\usepackage{amsmath}	
\usepackage{amssymb}	
\usepackage{mathtools}
\usepackage{xcolor}
\definecolor{darkblue}{RGB}{0,20,125}
\hypersetup{colorlinks=true,linkcolor=darkblue,citecolor=darkblue,filecolor=darkblue,urlcolor=darkblue,}

\usepackage{newtxtext,newtxmath}

\renewcommand{\d}{\mathrm{d}}
\renewcommand{\th}{\vartheta}
\newcommand{\ph}{\varphi}
\newcommand{\cE}{\mathcal{E}}
\newcommand{\eps}{\varepsilon}
\newcommand{\ssq}{\sin^2\!}
\newcommand{\csq}{\cos^2\!}
\newcommand{\rc}{r_{\rm c}}
\newcommand{\Js}{J_{r \mathrm{s}}}

\title[Action-angle coordinates for black-hole geodesics I]{Action-angle coordinates for black-hole geodesics I: Spherically symmetric and Schwarzschild.}

\author[V. Witzany]{
V. Witzany,$^{1}$\thanks{E-mail: vojtech.witzany@ucd.ie}
\\
$^{1}$School of Mathematics and Statistics, University College Dublin, Belfield, Dublin 4, Ireland
}

\date{}

\pubyear{2021}

\begin{document}
\label{firstpage}
\pagerange{\pageref{firstpage}--\pageref{lastpage}}
\maketitle

\begin{abstract}
Action-angle coordinates are a tool commonly used in celestial mechanics to systematically parametrize and store general solutions of the equations of motion of astrophysical bodies. I perturbatively construct action-angle coordinates for bound test particle motion in static, spherically symmetric space-times using a post-circular expansion. Then I specialise the expressions to the motion in the gravitational fields of Schwarzschild black holes and give explicit formulas for the Hamiltonian to the 10th power in the radial action (20th power in eccentricity), and the transformation to angle coordinates up to the 8th harmonic with respect to a relativistic orbital anomaly. The results provide a closed-form perturbative solution for the orbital motion parametrized by coordinate time that will find applications in the modelling of compact binary inspirals and other fields of astrophysics. 
\end{abstract}




\section{Introduction}

In 1609 Johannes Kepler published the famous treatise \textit{Astronomia Nova} presenting for the first time his celebrated model of planetary motion. Already in that very treatise Kepler also introduced the notion of a ``mean anomaly'', an abstract phase variable which marched steadily forward through time while parametrizing the cycles of the closed orbits of the planets \citep{stephenson1994kepler}. Such phases since then became one of the cornerstones of various averaging and perturbation methods used in celestial mechanics and dynamical astronomy \citep[see, e.g.,][]{morbidelli2002modern,contopoulos2002order}. 

In modern Hamiltonian mechanics we know such phases as the ``angles'' in the canonical action-angle (AA) coordinates in phase space. In short, when we have an integrable autonomous Hamiltonian system of $N$ degrees of freedom fulfilling certain non-degeneracy and smoothness requirements, we can always cover its phase space with $N$ pairs of canonical coordinates $\psi^a,J_a, a=1,...,N$ such that the Hamiltonian is only a function of the ``actions'' $J_a$, $H = H(J_a)$. The corresponding equations of motion then read
\begin{align}
    \dot{J}_a = \frac{\partial H}{\partial \psi^a} = 0 \,,\; \dot{\psi}^a = \frac{\partial H}{\partial J_a} = \rm constant.
\end{align}
Since the angles $\psi^a$ march homogeneously forward in time, we see how they are a natural generalization of the mean anomaly of Kepler. As a result, by obtaining the transformation between AA coordinates and the original phase-space coordinates, one essentially obtains the general solution to the evolution equations. Additionally, when the system is perturbed, the AA coordinates form an elegant and efficient basis for the analytical computation of corrections to the system's evolution \citep{arnol2013mathematical,morbidelli2002modern,kevorkian2012multiple}.

Even though the analytical methods of perturbation theory have brought about a number of advances in our understanding of the solar system and beyond, they are now often replaced by much simpler numerical integration in computers, at least when precision beyond back-of-the-envelope estimates is needed. The argument for this approach is that, even though more computational power is typically needed in the numerical approach, one saves time on deriving and implementing the complicated analytical formulas in return. So the question stands: Is it still worth to develop advanced analytical approximations in this day and age? 

Amongst other examples, analytical perturbation methods brought to light, through the meticulous computations of \citet{1859AnPar...5....1L}, the discrepancy between Newtonian gravity and observation in the motion of Mercury and, coincidentally, gave one of the first confirmations of Einstein's general relativity \citep{1915SPAW...47..831E}. Since then, Einstein's gravity has matured into a theory used routinely in observations, with gravitational-wave astronomy being only one of the most recent examples. After the spectacular successes of the first three runs of LIGO and Virgo \citep{LIGOScientific:2018mvr,LIGOScientific:2020ibl,LIGOScientific:2021djp}, we are now also preparing for the launch of the gravitational-wave observatory Laser Interferometer Space Antenna (LISA) in the 2030s \citep[][]{amaro2017laser}. One of the key sources for LISA will be inspirals of stellar-mass compact objects into massive black holes known as extreme mass ratio inspirals \citep{Babak:2017tow}. These systems are most efficiently described by iterative perturbations away from geodesic motion in the massive black-hole background  \citep[see][for recent reviews]{Barack:2018yvs,Pound:2021qin}. Should we return to analytical perturbation methods in this case? 

It turns out that the matched-filtering methods of LISA data analysis will require the fast evaluation of an extremely large number of waveforms for various parameter choices of the gravitational-wave sources. Naively, for $P$ parameters of any source one will require an order of $10^P$ evaluations of the model, where $P$ is 7 to 14 for extreme mass ratio inspirals, dependent on the parameter counting \citep{Gair:2004iv}. In that case almost almost any fractional speedup of computations will be worth the time spent deriving and implementing analytical formulas ``offline''. This point is further strengthened by the fact that much of the analytical computations can now be relegated to symbolic software such as \textit{SageMath}, \textit{Mathematica}, or \textit{Maple}, and the same software can also automatically output code implementing the analytical formulas in lower-level programming languages such as C. Thus, it seems that General relativity is returning the service to analytical perturbation methods by reviving their relevance through gravitational-wave science.

In this paper, I analytically derive approximate AA coordinates for the trajectories of bound free test particles (geodesics) parametrized by coordinate time in any static, spherically symmetric metric field, and in particular the field of a non-rotating black hole in Einstein's relativity (Schwarzschild space-time). The AA coordinates for geodesics in the fields of spinning black holes (Kerr space-time) will be the subject of the second paper in this series. 

Black hole geodesics are important because, as already mentioned, they play the role of a zeroth-order system in extreme mass ratio inspirals. With the AA coordinates, one is then able to absorb the non-geodesic terms in the evolution equations of the inspiral and obtain an extremely efficient evolution scheme (for more, see \citet[][]{Hinderer:2008dm,VanDeMeent:2018cgn,Miller:2020bft,Pound:2021qin} and the recent applications in \citet{McCart:2021upc,Lynch:2021ogr}). Beyond that, AA coordinates provide an avenue for coordinate-independent comparisons between approaches to the relativistic two-body problem \citep{LeTiec:2011ab,LeTiec:2015kgg, Fujita:2016igj}. 

This work is related to previous works that derived AA coordinates under Carter-Mino parametrization in terms of integral formulas and special functions in Kerr space-time \citep{Schmidt:2002qk,Fujita:2009bp, vandeMeent:2019cam}. However, here the goal is to obtain AA coordinates under the \textit{coordinate-time} parametrization (time of faraway observers) in fully closed form. 

The idea of the herein presented derivation, which can be applied in any static spherically symmetric metric with stable circular orbits, is to compute the AA variables perturbatively with circular orbits taken as the zeroth-order reference. This approach is presented in Section \ref{sec:spher} with the details for general static spherically symmetric metrics given in Appendix \ref{app:AAspher}. In Section \ref{sec:schw} I then treat AA coordinates for the specific case of Schwarzschild space-time and use the parametrization due to \citet{darwin1961gravity} to increase the achieved order of expansion. Finally, the convergence and accuracy of the Schwarzschild AA coordinates is discussed in Section \ref{sec:conv}. 

This paper was written in parallel with the work of \citet{Polcar:2022quad} that carries out some analogous derivations; the correspondence between the works is discussed in the last Section \ref{sec:discuss}. 

{\em Supplemental material:} It would not be practical to print all the long formulas in this paper, so they are instead published in machine-readable form along with the \textit{Mathematica} notebooks used for the derivation as an online Supplemental material to this paper. All these materials are also published at \texttt{\href{https://www.github.com/VojtechW/Action-Angle-Schwarzschild}{github.com/VojtechW/Action-Angle-Schwarzschild}}.

{\em Notation:} $G=c=1$ units and Einstein summation are used throughout this paper. Greek letters $\mu,\nu$ as indices run from 0 to 4.

\section{AA coordinates in static, spherically symmetric space-times}
\label{sec:spher}

Static, spherically symmetric space-times represent an important idealized class of space-times that contains, amongst other examples, the gravitational fields of spherically symmetric black holes in various relativistic theories of gravity. Here I present a general treatment of the perturbative construction of action-angle coordinates in these space-times in a post-circular expansion, keeping in mind particularly the application to Schwarzschild space-time to be discussed in Section \ref{sec:schw}.

\subsection{Geodesics in spherically symmetric space-times}
\label{subsec:spher}
Static, spherically symmetric space-times are space-times with 4 isometries expressed by the existence of 4 Killing vectors $\xi_{(t)}^\mu, \xi_{(x)}^\mu, \xi_{(y)}^\mu, \xi_{(z)}^\mu$ corresponding to infinitesimal time translations and 3 spatial rotations respectively. In general, one can then pick spherical-polar coordinates $t,R,\vartheta,\varphi$ such that the metric attains the form
\begin{align}
    \d s^2 = - a(R) \d t^2 + b(R) \d R^2 + c(R) (\ssq \th \d \ph^2 + \d \th^2)\,, \label{eq:genmet}
\end{align}
where $a,b,c$ are some set of metric functions. Note that this form is sufficiently flexible to capture most commonly used sets of coordinates covering black-hole space-times outside of horizons, such as isotropic, harmonic, tortoise, or Schwarzschild-type coordinates.

Now the motion of massive test particles in these space-times is generated by the Hamiltonian (compare, e.g., eq. (9) in \citet{Witzany:2016jqv})
\begin{align}
\begin{split}
    H_t & = \sqrt{-\frac{g^{RR}p_R^2 + g^{\th\th}p_\th^2 + g^{\ph\ph} p_\ph^2 + 1}{g^{tt}}} 
    \\
    &= \sqrt{\frac{a}{b}p_R^2 + \frac{a}{c} \left (p_\th^2 + \frac{p_\ph^2}{\ssq \th} \right) + a} \,, \label{eq:Ht}
\end{split}
\end{align}
where $p_R,p_\th,p_\ph$ are canonically conjugate to $R,\th,\ph$.
This Hamiltonian is conserved with a constant value $\cE \equiv H_t$ with the meaning of orbital energy per unit mass of the particle. Additional constants of motion form an angular-momentum vector per unit mass defined as $\vec{\ell}\equiv (p_\mu \xi^\mu_{(x)},p_\mu \xi^\mu_{(y)},p_\mu \xi^\mu_{(z)})$. Only two integrals in involution can be constructed from  the angular-momentum vector, these can be chosen as:
\begin{align}
    & \ell_{(z)} \equiv p_\mu \xi^\mu_{(z)} =  p_\ph \,,\\
    & \ell \equiv \sqrt{p_\th^2 + \frac{p_\ph^2}{\ssq \th}}\,. \label{eq:ell}
\end{align}
Here $\ell_{(z)}$ is the azimuthal angular momentum per particle mass (the axis around which the $\ph$ rotates is conventionally the $z$-axis), and $\ell$ is the total magnitude of angular momentum per unit mass.

\subsection{Hamilton-Jacobi equation and separation of variables}
\label{subsec:HJ}

The generating function for the canonical transformation to action-angle variables is the Hamilton-Jacobi action $S(x^\mu, C_i)$, where $C_i$ is a set of separation constants (see, e.g., \citet{arnol2013mathematical}). The action fulfills the Hamilton-Jacobi equation
\begin{align}
    \frac{\partial S}{\partial t} = \sqrt{\frac{a}{b} \left(\frac{\partial S}{\partial R}\right)^2 + \frac{a}{c} \left(\left(\frac{\partial S}{\partial \th}\right)^2 + \frac{1}{\ssq \th}\left(\frac{\partial S}{\partial\ph}\right)^2 +a\right)}\,.
\end{align}
A solution can be easily obtained in separable form by substituting
\begin{align}
    S = \cE t + \ell_{\rm(z)} \ph + S_{(R)} (R) + S_{(\th)}(\th)\,.
\end{align}
Then we obtain
\begin{align}
    \frac{c (\cE^2-a)}{a} - \frac{c}{b} \left(S'_{(R)}\right)^{2} = \left(S'_{(\th)} \right)^{2} + \frac{\ell_{(z)}^2}{\ssq \th}\,.
\end{align}
We see that the left-hand side is only a function of $R$, and the right-hand side is only a function of $\th$, so they both must be proportional to a separation constant. By comparing with equation \eqref{eq:ell}, we see that the separation constant has to be equal to $\ell^2$, so we have two equations for the functions in the separable Ansatz for $S$
\begin{align}
    & S_{(R)} = \pm\int \! \sqrt{A(R)\left[\cE^2 - B(R;\ell)\right]} \d R\,, \label{eq:sr}\\
    & S_{(\th)} = \pm\int \! \sqrt{\frac{\ell^2 \ssq \th -\ell_{(z)}^2}{\ssq\th}} \d \th\,, \label{eq:sth}\\
    & A(R) \equiv  \frac{b}{a}\,,\; B(R;\ell) \equiv b \left(\frac{\ell^2}{c} + 1 \right)\,.
\end{align}
Note that there is a sign ambiguity in the solution, since we only obtain equations for the squares of the derivatives of $S_{(R)},S_{(\th)}$. 

The indefinite integral for $S_{(R)}$ cannot be found in closed form for more complicated metrics. On the other hand, the $S_{(\th)}$ integral can be expressed up to an integration constant as 
\begin{align}
\begin{split}
  \pm S_{(\th)} =& \, \ell_{(z)} \arctan\left( \frac{\ell_{(z)} \cos \th}{\sqrt{\ell^2 \ssq\th - \ell_{(z)}^2}}\right) 
  \\
  & +  \ell \arctan\left( \frac{\sqrt{\ell^2 \ssq\th - \ell_{(z)}^2}}{\ell  \cos \th}\right) \,.
\end{split}
\end{align}
It is desirable to choose different branches of $\arctan(x)$ for each of the terms so that the action is smooth around $\th=\pi/2$. Specifically, the branch in the first term should be such that $\arctan(0^+) = \arctan(0^-)$, and the second term should have $\arctan(+\infty) = \arctan(-\infty)$ (see also \citet{Polcar:2019kwu} for an alternative expression).

\subsection{Definition of actions}
\label{subsec:Acdef}

Given the solution of the Hamilton-Jacobi equation, I can now easily generate actions as functions of separation constants (integrals of motion) by defining them as integrals of momenta over independent loops in phase space \citep{arnol2013mathematical}. In this case I define the actions by phase-space integrals over independent librations in $R,\th$, and a $\ph$-rotation  to obtain
\begin{align}
    & J_R = \frac{1}{2 \pi} \! \oint\!\! p_R \d R = \frac{1}{\pi} \int_{R_1}^{R_2} \! \sqrt{A(R)\left[\cE^2 - B(R;\ell)\right]} \d R\,, \label{eq:jr}\\
    & J_{\th} = \frac{1}{2 \pi} \! \oint \!\! p_\th \d \th = \frac{2}{\pi} \int_{\pi/2}^{\th_{\rm max}} \! \sqrt{\frac{\ell^2 \ssq \th -\ell_{(z)}^2}{\ssq\th}} \d \th\,, \label{eq:jth}\\
    & J_\varphi = \frac{1}{2 \pi} \! \oint \!\! p_\ph \d \ph = \ell_{(z)} \label{eq:jph}\,,
\end{align}
where $R_1,R_2$ denote the radial turning points of the motion, and $\th_{\rm max} = \arcsin(\ell_{(z)}/\ell)$ the turning point in $\th$-coordinate space. While $J_R$ typically has no closed form, the $J_\vartheta$ variable can be shown to be equal to
\begin{align}
    J_{\vartheta} = \ell - |\ell_{(z)}|\,.
\end{align}
Note that the definitions of $J_R,J_\th,J_\ph$ are topological, so they will not change by switching to different sets of coordinates (as long as homotopically equivalent loops are chosen for them in every coordinate system). In particular, the value of $J_R$ for a given orbit is not sensitive to the choice of the radial coordinate.

\subsection{Obtaining angles}
\label{subsec:AAtrafo}
The expression of the angle coordinates canonically conjugate to the actions can be obtained by re-expressing the action $S$ in terms of $J_i = J_R,J_\th,J_\ph$ instead of constants of motion $C_j =\cE,\ell_{(z)},\ell$. Then one can take the transformed action $S(x^\mu,J_i) = S(x^\mu,C_j(J_i))$ and define the transformation to angle variables as 
\begin{align}
    \psi^i(x^\mu, J_k) = \frac{\partial S}{\partial J_i}(x^\mu, J_k) \,. \label{eq:psigen}
\end{align}
Interestingly, in many contexts it is more advantageous to carry out an \textit{indirect} procedure with the resulting angles still expressed as a function of constants of motion, $\psi^i = \psi^i(x^\mu, C_j)$ \citep{Schmidt:2002qk}. One starts by computing the Jacobian matrix of the transform given in equations \eqref{eq:jr}-\eqref{eq:jph} $\partial J_i/\partial C_j (C)$, finds its inverse matrix $\partial C_j/\partial J_i (C_k)$, and then one can define the transform
\begin{align}
    \psi^i = \frac{\partial S}{\partial J_i} = \frac{\partial S}{\partial C_j} (x^\mu, C_k) \frac{\partial C_j}{\partial J_i} (C_k)\,. \label{eq:indir}
\end{align}
The point of this trick is that the final expression on the right-hand side refers only to the separation constants $C_j$ and one does not need to necessarily invert the relation $J_i(C_j)$. 

When the dust settles, equation \eqref{eq:indir} yields in our case
\begin{align}
    & \psi^R = \frac{1}{\partial J_R/\partial \cE} \frac{\partial S}{\partial \cE}\,, \label{eq:psir} \\
    & \psi^\th = \frac{\partial S}{\partial \ell} -\frac{\partial J_R/\partial \ell}{\partial J_R/\partial \cE} \frac{\partial S}{\partial \cE} \,, \label{eq:psith} \\
    & \psi^\ph = \frac{\partial S}{\partial \ell_{(z)}} + \mathrm{sign}(\ell_{(z)}) \left(\frac{\partial S}{\partial \ell}  -\frac{\partial J_R/\partial \ell}{\partial J_R/\partial \cE} \frac{\partial S}{\partial \cE} \right) \label{eq:psiph}\,,
\end{align}
where the partial derivatives are always computed while keeping the other members of the variable set $x^\mu, \cE,\ell,\ell_{(z)}$ constant. 

It is also useful to note what do the partial-derivatives of $J_r$ represent when expressed in terms of the action-angle Hamiltonian $\cE(J_R,J_\varphi,J_\vartheta)$. In fact, the Hamiltonian will always depend only on $\ell = J_\vartheta + |J_\varphi|$ due to the spherical symmetry, so we can write
\begin{align}
    & \left(\frac{\partial J_R}{\partial \cE}\Big|_{\ell = \rm const.}\right)^{-1} = \frac{\partial \cE}{\partial J_R}\Big|_{\ell = \rm const.} = \Omega^R \,, \\
    & \frac{\partial J_R/\partial \ell|_{\cE = \rm const.}}{\partial J_R/\partial \cE|_{\ell = \rm const.}} = -\frac{\partial \cE}{\partial \ell}|_{J_r = \rm const.} = -\Omega^\th = -{\rm sign}(J_\ph) \Omega^\ph\,, 
\end{align}
where $\Omega^i \equiv \dot{\psi}^i = \partial \cE/\partial J_i,\,i=R,\th,\ph$ are the fundamental frequencies of motion with respect to coordinate time $t$. It should be kept in mind that different time parametrizations lead to the same actions, but generally to different angle coordinates; this set is tied to the $t$-parametrization. The cases of parametrization by other parameters such as proper time or Carter-Mino time are discussed in Appendix \ref{app:param}.

\subsection{Delaunay-type variables} \label{subsec:delaunay}
Variables analogous to Delaunay variables can be obtained by choosing coordinates (cf. \citet{morbidelli2002modern})
\begin{align}
\begin{split}
& J_{1} \equiv J_r + J_{\th} +|J_\ph|\,,\\
& J_{2} \equiv \ell = J_{\th} +|J_\ph| \,,\\
& J_3 \equiv \ell_{\rm(z)} = J_\ph\,,
\end{split}
\begin{split}
& \psi^1 = \psi^r \,,\\
& \psi^2 =  \psi^\th-\psi^r\,,\\
& \psi^3 =   \psi^\ph-{\rm sign}(J_\ph)\psi^\th\,.
\end{split}
\end{align}
In the Newtonian problem of two point masses the Hamiltonian ends up being dependent only on $J_1$. However, note that for strong-field Schwarzschild geodesics the Hamiltonian is a non-trivial function of both $J_1,J_2$. As in the Kepler problem, the variable $\psi^1$ can be assigned the loose meaning of the mean anomaly, $\psi^2$ the argument of periapsis, and $\psi^3$ can be understood as the longitude of ascending node. The angle $\psi^2$ is not constant in relativity (which corresponds to pericenter precession), but the longitude of ascending node $\psi^3$ is constant in spherically symmetric space-times (which corresponds to a fixed orbital plane).

\subsection{Post-circular expansion of AA coordinates}

 At this point I assume that there is at least one family of stable circular orbits in the space-time in question. Mathematically, the conditions for this to be true boil down to the requirement that there exists a one-parameter family of solutions to the equations
 \begin{align}
     & \cE^2 - B(R;\ell)  =0 \,, \label{eq:cfirst}\\
     & \frac{\d}{\d R} \left( \cE^2 - B(R;\ell) \right) = 0\,,\\
     & \frac{\d^2}{\d R^2} \left( \cE^2 - B(R;\ell) \right) > 0 \label{eq:clast}\,,
 \end{align}
 where $ \ell, \cE, R$ are the variables to be solved for. I also assume that we can parametrize the circular orbits by $\ell$. That is, I assume that for $\ell$ in some range, the equations \eqref{eq:cfirst}-\eqref{eq:clast} are solved by $ \ell, \cE_{\rm c}(\ell), R_{\rm c}(\ell)$. 
 
 The idea of the post-circular expansion is to expand both the integrands and the integral bounds $R_1,R_2$ in equations \eqref{eq:sr} and \eqref{eq:jr} in terms of $\Delta \cE = \cE - \cE_{\rm c}(\ell)$. Since at $\cE = \cE_{\rm c}(\ell)$ we have $J_R = 0$, one can also view this as an expansion in $J_R$. 
 
 Finally, since this causes the orbit to be non-circular, it can also be seen as an expansion in a parameter proportional to some notion of eccentricity $e$, since $e \sim \sqrt{\Delta \cE} \sim \sqrt{J_R}$. Do note, however, that definitions of $e$ are non-unique (coordinate-dependent) in relativity, and that one should expect a different convergence behaviour of the expansion as compared to a true eccentricity expansion, since $J_R \to \infty$ as eccentricity goes to $1$. Specifically, the $J_R$ power expansions allow to  automatically include divergences that scale as $1/(1-e)^\alpha$ by naturally including terms that scale as $J_R^{\alpha/2}$.
 
 The real focus of this paper is Schwarzschild space-time, so I relegate the technical details of the expansion in general spherically symmetric space-times to Appendix \ref{app:AAspher}. At this point it suffices to say that in full generality the Hamiltonian is expanded as 
 \begin{align}
    & \cE = \cE_{\rm c}(\ell) + \sum_{i=1}^{[n/2]}J_R^i \eps_i(\ell)\,, \label{eq:EfinM}\\
    & \eps_1 = \sqrt{\frac{B^{''}}{2 \cE_{\rm c}^2 A}}\,,\; \eps_2 = \frac{\cE_{\rm c}^2 \mathcal{F}  - 12A^2 B^{''3}}{48 \cE_{\rm c}^3 A^3 B^{''2}}\,,\\
    \begin{split}
     & \mathcal{F} \equiv 3A^{'2} B^{''2} + 6A\left(A^{'} B^{''} B^{(3)} -A^{''}B^{''2}\right) \\  & \phantom{\mathcal{F} \equiv} + A^2 \left(3B^{''} B^{(4)} - 5(B^{(3)})^2\right) \,,
    \end{split} \\
    & \eps_3 = ... \nonumber\,,
\end{align}
where expressions up to $\varepsilon_5$ in any static spherically symmetric space-times can be found in the supplemental Mathematica data files. Note that the expressions are purposefully parametrized by $R_{\rm c}, \cE_{\rm c}$ while assuming that the expression for $R_{\rm c}(\ell), \cE_{\rm c}(\ell)$ will be substituted into the final result. Now one can see that upon the substitution $\ell = J_\th + |J_\ph|$ into \eqref{eq:EfinM}  we obtain a perturbative expression for the Hamiltonian $H_t$ in terms of AA coordinates, $H_t = \cE(J_R,J_\th,J_\ph)$. It is also possible to obtain the full transformation to angle variables and to express fundamental frequencies of motion, as also described in Appendix \ref{app:AAspher} and one of the supplemental notebooks.

\section{AA coordinates in Schwarzschild space-time} \label{sec:schw}

The Schwarzschild metric in Schwarzschild coordinates $t,r,\th,\ph$  reads
\begin{align}
    \d s^2 = -\left(1 - \frac{2 M}{r}\right) \d t^2  \frac{1}{1-2M/r} \d r^2 + r^2 \left(\ssq\th \d \ph^2 + \d \th^2 \right).
\end{align}
Obviously, this metric is in the form \eqref{eq:genmet} with $R\to r$ and $c=r^2,\,a= 1/b = 1-2M/r$. In this section I compute the transformation to AA coordinates from the Schwarzschild-coordinate basis. Note that using a different radial coordinate than $r$ only changes the transformation to the angle variables and leaves the actions intact.

The expansions can be obtained by substituting the following auxiliary functions and relations for stable circular orbits into the coefficients from Section \ref{sec:spher} and Appendix \ref{app:AAspher}:
\begin{align}
    & A(r) = \left( 1 - \frac{2 M}{r}\right)^{-2}\!,\\
    & B(r;\ell) = \left(1 + \frac{\ell^2}{r^2} \right) \left(1 - \frac{2 M}{r} \right),\\
    & C(r) = \frac{1}{r^2}\,,\\
    & r_{\rm c}(\ell) = \frac{\ell}{2 M} \left(\ell + \sqrt{\ell^2 - 12 M^2} \right)\,,\\
    \begin{split}
    & \cE_{\rm c}(\ell) = \frac{r_{\rm c}- 2M}{\sqrt{r_{\rm c}(r_{\rm c} - 3M)}}
    \\ & \phantom{\cE_{\rm c}(\ell)} = \sqrt{\frac{2}{3} + \frac{2\sqrt{\ell^2 - 12 M^2}}{9 \ell} + \frac{\ell (\ell - \sqrt{\ell^2 - 12M^2})}{54 M^2}} \,.
    \end{split}
\end{align}
Stable circular orbits exist only for $\ell> \sqrt{12}M$ ($r_{\rm c}(\ell)>6 M$) and the formalism is generally expected to become ill-convergent already for $\ell \sim \sqrt{12}M$. Another important transitional point is $\ell = 4M$. Above this angular momentum, the transition from bound to scattering motion (motion escaping to infinity) occurs at $\cE = 1$. However, for $\ell<4M$ the orbits stop oscillating around the stable circular orbit already at the energy of the so-called homoclinic orbit (infinite zoom-whirl orbit)
\begin{align}
    \cE_{\rm h} = \sqrt{\frac{2}{3} - \frac{2\sqrt{\ell^2 - 12 M^2}}{9 \ell} + \frac{\ell (\ell + \sqrt{\ell^2 - 12M^2})}{54 M^2}} \,, \label{eq:Eh}
\end{align}
and the transition is to orbits plunging into the black hole horizon, $r\to 0$. The values $\cE =1, \cE = \cE_{\rm h}$ for $\ell>4M$ and $\sqrt{12}M<\ell<4M$ respectively represent critical energies near which one should expect the post-circular expansion to diverge at given $\ell$, unless treated by Padé resummation or a similar technique (see Section \ref{sec:conv} and Appendix \ref{app:asymp} for more).

 It turns out that in Schwarzschild space-time one can push the expansions to much higher order by parametrizing by Darwin's semi-latus rectum and eccentricity $p,e$, and by keeping the position of the minimum of the effective potential $r_{\rm c}(\ell)$ implicit until evaluation. I now describe this expansion as tailored specifically to the Schwarzschild metric in Schwarzschild coordinates.

\subsection{Darwin's eccentricity and semilatus rectum} \label{subsec:darwin}
Instead of the integrals of motion $\cE,\ell$ one can choose to parametrize the orbits by their turning points $r_1<r_2$. Specifically, \citet{darwin1961gravity} introduced the analogue of the Newtonian eccentricity $e$ and semi-latus rectum $p$ in Schwarzschild coordinates as follows
\begin{align}
    &r_1 = \frac{p}{1 + e}\,,\; r_2 = \frac{p}{1 - e} \,,\\
    \Rightarrow \, & e \equiv \frac{r_2 - r_1}{r_1 + r_2}\,,\; p \equiv \frac{2 r_1 r_2}{r_1 - r_2}\,. \label{eq:epr}
\end{align}
Energy and angular momentum can then be given in terms of these as
\begin{align}
    & \cE^2 = \frac{(p - 2M)^2 - 4M^2 e^2}{p\left[p - M(3+e^2)\right]}\,,\; \ell^2 = \frac{M p^2}{p - M(3+e^2)}\,. \label{eq:Epe}
\end{align}
A useful property of the $p,e$ parametrization is also that the separatrix between stable bound motion and plunge into the black hole is given by the simple formula $p = (6+2e)M$ \citep{Cutler:1994pb,Stein:2019buj}.

The $p-e$ formalism can also be used in the parametrization the ``anomaly'' of the radial motion. By a method of trial and error I found that the $\xi$-parametrization of the radial motion introduced in Appendix \ref{app:AAspher} leads to unmanageable expressions at higher expansion order in Schwarzschild space-time. It turns out that high order in $J_r$ can instead be reached by using the relativistic analogue of the Keplerian true anomaly, known as the \textit{relativistic anomaly} $v$ \citep{darwin1961gravity,Schmidt:2002qk}:
\begin{align}
    r = \frac{p}{1 + e \cos v} \,.
\end{align}
The action integrals \eqref{eq:sr} and \eqref{eq:jr} are then expressed as
\begin{align}
\begin{split}
    & S_{(r)} =\int_0^{v(r)} \!\!\Bigg[\sqrt{\frac{p-6M - 2 e M\cos v'}{p-M(3 + e^2)}}\times\\
    & \phantom{S_{(r)} =\int_0^{v(r)} \!\!\Bigg[} \frac{M^{1/2} e^2 p^{3/2} \sin ^2\!v'}{(p-2M +2e M \cos v')(1 + e \cos v')^2}\Bigg] \d v',  \label{eq:srep}
\end{split} 
\\ & J_r = \frac{1}{\uppi}S_{(r)}\Big|_{v(r)=\uppi}. \label{eq:Jrep}
\end{align}
The $S_{(r)}$ integral is expressible as a complicated expression involving elliptic functions, but here I will express it using the post-circular expansion, since this would be either way necessary for the full set of transformations.

To expand these expressions in terms of eccentricity, I first re-express $p$ as $p = p(r_{\rm c}(\ell),e)$
\begin{align}
    p = \frac{r_{\rm c}\left[r_{\rm c} + \sqrt{r_{\rm c} - 4(r_{\rm c} - 3M)(3+e^2)}\right]}{2(r_{\rm c} - 3M)}\,. \label{eq:prc}
\end{align}
This can also be inverted to obtain
\begin{align}
    \rc = \frac{p^2 + p \sqrt{(p-6M)^2 +12M^2 e^2}}{2(p-3M-M^2e^2)}\,.
\end{align}
The reason for using $r_{\rm c}(\ell)$ is that the expansions generally stay more compact than when using $\ell$ as a variable directly. However, the relation between $J_r,\rc(\ell)$ and $p,e$ cannot be expressed entirely in closed form. One can expand the integral \eqref{eq:Jrep} in powers of $e$ as
\begin{align}
    & J_r = \sum_{i = 1}^{[n/2]} \tilde{j}_{ri}(p) \kappa^{2i} e^{2i}\,, \\
    & \tilde{j}_{r1} = \frac{p^{3/2}}{2(p - 2M)} \sqrt{\frac{p-6M}{p - 2M}}\,,\;\tilde{j}_{r2} = ... \,,
\end{align}
Then one can substitute equation \eqref{eq:prc} for $p(r_{\rm c},e)$, reexpand in $e$ and invert the perturbative relation to obtain
\begin{align}
\Rightarrow \, & e^2 = \sum_{i = 1}^{[n/2]} \kappa^{2i}\mathfrak{E}_i(\rc) J_r^i \,, \label{eq:ecJexp}\\
    & \mathfrak{E}_1 = \frac{2(r_{\rm c} - 2M)}{r_{\rm c}^{3/2}} \sqrt{\frac{r_{\rm c} - 2M}{r_{\rm c}-6M}}\,,\;\mathfrak{E}_2 = ... \,.
\end{align}
 Using these formulas, it is possible to perturbatively compute $J_r,\rc$ given $e,p$ and vice versa. It is possible to easily reach order $n=20$ and higher in these expressions.

Plugging in the identity \eqref{eq:prc} and subsequently expansion \eqref{eq:ecJexp} into the exact relation for $\cE(e,p)$ given in equation \eqref{eq:Epe} I also obtained the Hamiltonian (energy) coefficients $\eps_i$ to $n=20$. The formulas up to $\varepsilon_5$ ($n=10,\,\mathcal{O}(J_r^5)$) are given in Table \ref{tab:ekoef}. Expansion coefficients up to $\varepsilon_{10},\mathfrak{E}_{10},\tilde{j}_{r10}$ are given in one of the supplemental Mathematica files.

\begin{table*}
\caption{Energy expansion coefficients in Schwarzschild space-time (see eq. \eqref{eq:EfinM}, $l\equiv \ell/M,\, \lambda \equiv \sqrt{l^2 - 12}$).}
\label{tab:ekoef}
\begin{tabular}{ll}
\hline
$M \eps_1(\ell)$ & $\frac{1}{l^2}\sqrt{\frac{2 \lambda}{l^3+\lambda  l^2-9 l-3 \lambda}}$ 
\vspace{0.5 em} \\
$M^2 \eps_2(\ell)$ & $\frac{(5 l^6-5 \lambda  l^5+161 l^4-299 \lambda  l^3-4440 l^2+4068 \lambda 
   l+21456)\sqrt{l^3-\lambda  l^2+36 l+12 \lambda}}{216 \sqrt{6} l^{7/2} \lambda^3 \left(l^2+4\right)}$ 
\vspace{0.5 em} \\
$M^3\eps_3(\ell)$ & $\frac{\sqrt{2 l^6+2 \lambda  l^5-42 l^4-30 \lambda  l^3+234 l^2+90 \lambda 
   l-216} \left(-235 l^{10}+235 \lambda  l^9-22560 l^8+23970 \lambda  l^7+1072800 l^6-924750 \lambda 
   l^5-15736140 l^4+10644480 \lambda  l^3+90940482 l^2-39680280 \lambda 
   l-154627704 \right)}{1259712 l^5 \lambda^5}$ 
\vspace{0.5 em} \\
$M^4\eps_4(\ell)$ & $\frac{5 \sqrt{l^3 - \lambda l^2 + 36 l +12 \lambda}}{5038848 \sqrt{6} l^{13/2} \lambda^7 \left(l^2+4\right)}
   \times
   \begin{matrix} \scriptstyle\big[ 
   -3559 l^{11}+3559 \lambda  l^{10}-1315465 l^9+497011 \lambda  l^8+77098428 l^7-35631084
   \lambda  l^6-1600484859 l^5
   & \\ \scriptstyle
   +793886589 \lambda  l^4+14410520136 l^3-7249631976
   \lambda  l^2-47679449904 l+23993658000 \lambda
   \big]
   \end{matrix}$  
   
\vspace{0.5 em} \\
$M^5\eps_5(\ell)$ & $\frac{320979616137216 \sqrt{6} \left(l^6+\lambda  l^5-19 l^4-13 \lambda  l^3+88 l^2+28 \lambda 
   l-48\right)^{3/2}}{(l^2+\lambda  l-12)^11 (l^2+\lambda  l-4)^6 (l + \lambda)^{35/2} l^8 (l^3-\lambda  l^2+36 l+12 \lambda)^{9/2}}
   \times
   \begin{matrix} \scriptstyle\big[ 
   1024 l^{48}+1024 \lambda  l^{47}-175616 l^{46}-169472 \lambda  l^{45}+14129024
   l^{44}+13130624 \lambda  l^{43}-708463232 l^{42}-632619392 \lambda 
   l^{41}
   & \\ \scriptstyle
   +24809065904 l^{40}+21232227248 \lambda  l^{39}-644343844144
   l^{38}-527049827344 \lambda  l^{37}+12866317595400 l^{36}
   & \\ \scriptstyle
   +10027421251080 \lambda 
   l^{35}-202065706764584 l^{34}-149527863111896 \lambda  l^{33}+2532392531036133
   l^{32}
   & \\ \scriptstyle
   +1772403298882437 \lambda  l^{31}
   -25550540213913464 l^{30}
   -16839160997188874
   \lambda  l^{29}+208478202242629456 l^{28}
   & \\ \scriptstyle
   +128733537628106782 \lambda 
   l^{27}
   -1376969800769007928 l^{26}
   -792073803428658604 \lambda 
   l^{25}+7345169113625041204 l^{24}
   & \\ \scriptstyle
   +3909803243910143746 \lambda 
   l^{23}
   -31471983075904562848 l^{22}
   -15381260683691871940 \lambda 
   l^{21}
   & \\ \scriptstyle
   +107363118990821829752 l^{20}
   +47731317526490643500 \lambda 
   l^{19}
   -287928198560779939944 l^{18}
   & \\ \scriptstyle
   -115145573742770984592 \lambda 
   l^{17}
   +596579918566652923690 l^{16}
   +211673522138430265984 \lambda 
   l^{15}
   & \\ \scriptstyle
   -932946796005105262272 l^{14}
   -288646890783487308288 \lambda 
   l^{13}
   +1066987636088749302144 l^{12}
   & \\ \scriptstyle
   +281468364609767964672 \lambda 
   l^{11}
   -854668190625260405760 l^{10}
   -186495412536706056192 \lambda 
   l^9
   & \\ \scriptstyle
   +450963625853655972864 l^8
   +77950463414486925312 \lambda 
   l^7
   -142981626878551080960 l^6-
   & \\ \scriptstyle
   18303137213746642944 \lambda  l^5
   +23421131332158849024
   l^4
   +1967525319633272832 \lambda  l^3
   -1476564694678634496 l^2
   & \\ \scriptstyle
   -60845365027405824
   \lambda  l+14880718205878272
   \big]
   \end{matrix}$  
\\
\hline
\end{tabular}
\end{table*}

\subsection{Orbital solutions} \label{subsec:orbsol}
The transformation to angle coordinates can in principle be obtained by expressing $S_{(r)} = S_{(r)}(J_r,v(r;J_r,r_{\rm c}(\ell)))$ in a trigonometric expansion and applying the partial derivatives according to eq. \eqref{eq:psigen} (including integral-bound terms due to non-zero $\partial v/\partial J_r,\partial v/\partial \ell$). However, I take the indirect route given in equations \eqref{eq:psir} to \eqref{eq:psiph} since one then avoids the integral-bound terms. The derivatives of the radial part of the action are given as
\begin{align}
\begin{split}
    & \frac{\partial S_{(r)}}{\partial \cE}\Big|_{\ell} = \int_0^{v(r)}\! \Bigg[ \frac{p^2}{(1+e\cos v')^2 \left[p - 2M(1+e\cos v')\right]} \times \\
    & \phantom{\frac{\partial S_{(r)}}{\partial \cE}|_{\ell} = \int_0^{v(r)} \Bigg[}  \sqrt{\frac{(p-2M)^2 - 4M^2 e^2}{p-M(6 + 2M e \cos v')}}\Bigg] \d v' \,,
\end{split}
    \\
    & \frac{\partial S_{(r)}}{\partial \ell}\Big|_{\cE} = -\int_0^{v(r)}\!\sqrt{\frac{p}{p - 2M(3 + e \cos v')}} \d v' \,,
\end{align}
The angle coordinates are then obtained by taking derivatives of the action according to equations \eqref{eq:psir} to \eqref{eq:psiph} (compare also section \ref{appsub:angles} in the Appendix). I start by taking a harmonic expansion of terms appearing in equations \eqref{eq:psith} and \eqref{eq:psiph}
\begin{align}
    & \frac{\partial S_{(r)}}{\partial \ell}\Big|_{\cE} - \frac{\partial J_r}{\partial \ell}\Big|_{\cE} \psi^r = \frac{\partial S_{(r)}}{\partial \ell}\Big|_{\cE} + \frac{\partial \cE}{\partial \ell}\Big|_{J_r} \frac{\partial S_{(r)}}{\partial \cE}\Big|_{\ell}
    \\& \equiv \sum_{j=1,k=1}^{n-1} \theta_{jk}(r_{\rm c}) J_{r}^{j/2} \sin(kv)\,, \label{eq:Sigr} 
\end{align}
where the quantities behind vertical lines denote variables that are kept fixed in the partial derivatives.
Note that the $\propto\psi^r$ term in the equation above exactly cancels the secularly growing term in $\partial S_{(r)}/\partial \ell$. As such, the resulting expression can be understood as the part of $\partial S_{(r)}/\partial \ell$ purely fluctuating with respect $\psi^r$. Also, no $\cos(k v)$ harmonics appear because of the reversibility of the motion and the choice of $v=0$ at the radial turning point. Then we have
\begin{align}
    & \psi^r = v + \sum_{j=1,k=1}^{n-1} \sigma_{jk}(r_{\rm c}) J_r^{j/2} \sin(k v)\,, \label{eq:psirv}\\
    & \psi^\th = \arctan\left(\frac{\sqrt{\ell^2 \ssq \th - \ell_{(z)}^2}}{\ell \cos\th}\right) +\!\! \sum_{j=1,k=1}^{n-1} \theta_{jk}(r_{\rm c}) J_{r}^{j/2} \sin(kv)\,, \label{eq:psithexp}\\
    \begin{split}
    & \psi^\ph = \ph + \arctan\left( \frac{\ell_{(z)} \cos \th}{\sqrt{\ell^2 \ssq \th - \ell_{(z)}^2}} \right) + \mathrm{sign}(\ell_{(z)}) \psi^\th \,. \label{eq:psiphexp}
    \end{split}
\end{align}
  The harmonic coefficients $\sigma_{jk}, \theta_{jk}$ fulfill 
 \begin{align}
     & \sigma_{jk} \neq 0,\, \theta_{ij} \neq 0\, \iff k \geq j,\, j+k \, {\rm even} , \\
     & \sigma_{11} = \frac{\sqrt{2(\rc^2 - 8 M \rc +12M^2)}(-2 \rc^2 + 19 M \rc -38 M^2    )  \mathcal{G}}{r_c(\rc - 6M)^2 (r_c - 2M)}, \label{eq:sig11}\\
    & \mathcal{G} =  \left[\frac{\rc}{M}(\rc^2 - 9 M r_c + 18 M^2)\right]^{1/4}, \\
    & \theta_{11} = - \frac{2 \sqrt{2} (\rc - 3M)^2}{\left[M\rc(\rc^2 - 9 M \rc +18 M^2)^3\right]^{1/4} \sqrt{\rc - 2M}}\,, \label{eq:th11} \\
    & \sigma_{22} = ...\,,\; \theta_{22} = ... \,.
 \end{align}
 Again, in equations \eqref{eq:psithexp} and \eqref{eq:psiphexp} different branches of $\arctan$ have to be used so that the expressions are regular at $\th = \pi/2$.
The harmonic expansion of $\psi^r$ can be perturbatively inverted to yield
\begin{align}
    & v(\psi^r) = \psi^r + \sum_{j=1,k=1}^{n-1} \Sigma_{jk}(r_{\rm c}) J_r^{j/2} \sin(k v)\,. \label{eq:vpsir}
\end{align}
Then it is convenient to solve for a new set of coefficients $\Theta_{ij}(r_{\rm c})$ from the equation
\begin{align}
\begin{split}
    & \!\sum_{j=1,k=1}^{n-1} \Theta_{jk}(r_{\rm c}) J_{r}^{j/2} \sin(k \psi^r)
    \\& = \sum_{j=1,k=1}^{n-1} \theta_{jk}(r_{\rm c}) J_{r}^{j/2} \sin\left(k v(\psi^r)\right) + \mathcal{O}(J_r^{n})\,.
\end{split}
\end{align}
The coefficients $\Theta_{jk}, \Sigma_{jk}$ share the same symmetries as $\theta_{jk},\sigma_{jk}$, that is, they are non-zero if and only if $k\geq j$ and $j+k$ is even. Their leading-order components fulfill $\Theta_{11} = \theta_{11}$ and $\Sigma_{11} = - \sigma_{11}$ with $\theta_{11}$ and $\sigma_{11}$ given in equations \eqref{eq:sig11} and \eqref{eq:th11}. Terms up to $n=8$ for all the harmonic coefficients are again given in the accompanying notebooks.

Now we can finally express $\th(\psi^r,\psi^\th),\, \ph(\psi^r,\psi^\th,\psi^\ph)$ as
\begin{align}
    & \cos \th = \sqrt{1 - \frac{\ell_{\rm (z)}^2}{\ell^2}} \, \cos\left(\psi^\th - \!\!\! \sum_{j=1,k=1}^{n-1} \Theta_{jk} J_{r}^{j/2} \sin(k \psi^r) \right),  \label{eq:thpsith}\\
    \begin{split}
    & \ph = \psi^\ph  - {\rm sign}(\ell_{\rm(z)}) \psi^\th
    \\ &\phantom{\ph = } - \arctan \left[\frac{\ell_{\rm(z)}}{\ell} \cot\left(\psi^\th -\!\!\! \sum_{j=1,k=1}^{n-1} \Theta_{jk} J_{r}^{j/2} \sin(k \psi^r)\right) \right]. \label{eq:phpsiph}
    \end{split}
\end{align}
The coefficients $\Theta_{jk}$ become quite complicated at higher order. Instead, one can avoid computing $\Theta_{ij}$ by evaluating $v(\psi^r)$ by using equation \eqref{eq:vpsir} and then using the $\theta_{jk}$ coefficients and $v$ harmonics for the fluctuating terms in the equations above.

\subsection{Fundamental frequencies of motion} \label{subsec:funfrek}

Since $\psi^r,\psi^\th,\psi^\ph$ evolve with uniform frequencies in coordinate time, equations \eqref{eq:vpsir}, \eqref{eq:thpsith} and \eqref{eq:phpsiph} can be understood as an analytical perturbative solution for the geodesic motion once the coefficients $\Sigma_{jk}, \Theta_{jk}$ are specified. The fundamental frequencies for the evolution of $\psi^r,\psi^\th,\psi^\ph$ are then obtained as
\begin{align}
    & \dot{\psi}^r \equiv \Omega^r = \frac{\partial H}{\partial J_r} = \sum_{i=1}^{[n/2]} i J_{R}^{i-1} \varepsilon_i(\rc)\,,\\
    \begin{split}
    & \dot{\psi}^\th \equiv \Omega^\th = \frac{\partial H}{\partial J_\th} = \frac{\d \cE_{\rm c}}{\d \ell}+ \sum_{i=1}^{[n/2]} \kappa^{2 i} J_{R}^{i} \frac{\d \varepsilon_i}{\d \ell}  \\
    & \phantom{\dot{\psi^\th}} = \frac{2 (\rc - 3M)^{3/2}}{\sqrt{M}(\rc - 6)}\left(\frac{\d \cE_{\rm c}}{\d \rc}+ \sum_{i=1}^{[n/2]} \kappa^{2 i} J_{R}^{i} \frac{\d \varepsilon_i}{\d \rc} \right) \,,
    \end{split} \\
    & \dot{\psi}^\ph \equiv \Omega^\ph = \frac{\partial H}{\partial J_\ph} = {\rm sign}(\ell_{(z)}) \Omega^\th,
\end{align}
where one can use that since the energy only depends on $\ell = J_\th + |J_\ph|$  and thus one can also replace $\partial/\partial J_\th \to \partial/\partial \ell$ and $\partial/\partial J_\ph \to {\rm sign}(\ell_{(z)}) \partial/\partial \ell$.

How are the frequencies $\Omega^r, \Omega^\th, \Omega^\ph$ related to real observable frequencies in $r,\th,\ph$? We see from equation \eqref{eq:psirv} that the geodesic finishes a full cycle in the radial motion during one cycle in the angle variable $\psi^R \in (0,2\pi]$. The frequency of radial motion is thus
\begin{align}
    f^r = \frac{\Omega^r}{2\pi}\,.
\end{align}
Now, as obvious from equations \eqref{eq:psith} and \eqref{eq:thpsith}, the $\th$ motion will not have a sharp period since it involves the fluctuating term $\sim \theta_{jk} \sim \Theta_{jk}$. However, its time-averaged frequency will be again identical to that of $\psi^\th$, or
\begin{align}
    \langle f^\th \rangle = \frac{\Omega^\th}{2\pi}\,.
\end{align}

Finally, there is the average rate of advance of the azimuthal angle $\ph$. The second and third terms on the right-hand side of eq. \eqref{eq:phpsiph} are purely fluctuating. The long-time average of $ \dot{\ph}$ is then actually given by $\dot{\psi}^\ph$ and we obtain
\begin{align}
    \langle f^\ph \rangle \equiv \frac{1}{2 \pi} \langle \dot{\ph} \rangle = \frac{\Omega^\ph}{2\pi} \,.
\end{align}

\section{Accuracy of Schwarzschild AA coordinates} \label{sec:conv}

\begin{figure}
    \centering
    \includegraphics{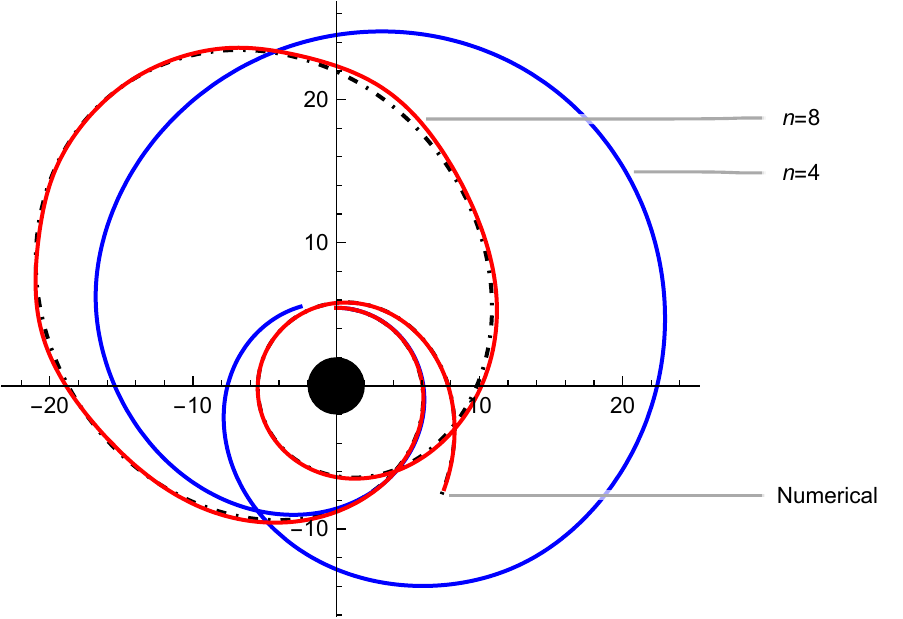}
    \caption{Analytical orbit with $e=0.6, p=10M$ in action angle coordinates (blue line: $n=4$, red line $n=8$) compared to a high-accuracy numerically integrated orbit (black dot-dashed). The $n=4$ expansion becomes very quickly inadequate for such a highly eccentric orbit near the limit of stability $p=(6+2e)M$, while the $n=8$ expansion still fares reasonably well. The axes correspond to $x= r\sin \ph\,,y= r\cos \ph$ plotted in units of $M$.}
    \label{fig:orb}
\end{figure}

The expressions obtained in Section \ref{sec:schw} can be understood as a Taylor series for the action-angle coordinates and for the related transformations expanded around $J_r = 0$ with respect to powers of $\sqrt{J_r}$. However, to which order is it sufficient to expand in practice? Amongst other thing, this depends on the position in orbital space. Figure \ref{fig:orb} illustrates that the higher-order expansions are able to provide satisfactory behaviour even for orbits quite close to the separatrix $p=(6+2e)M$. More precise statements can be given when a more concrete application of the herein presented formalism is considered. 

\subsection{Accuracy of Hamiltonian and resummation}

\begin{figure*}
    \centering
    \includegraphics[width=0.32\textwidth]{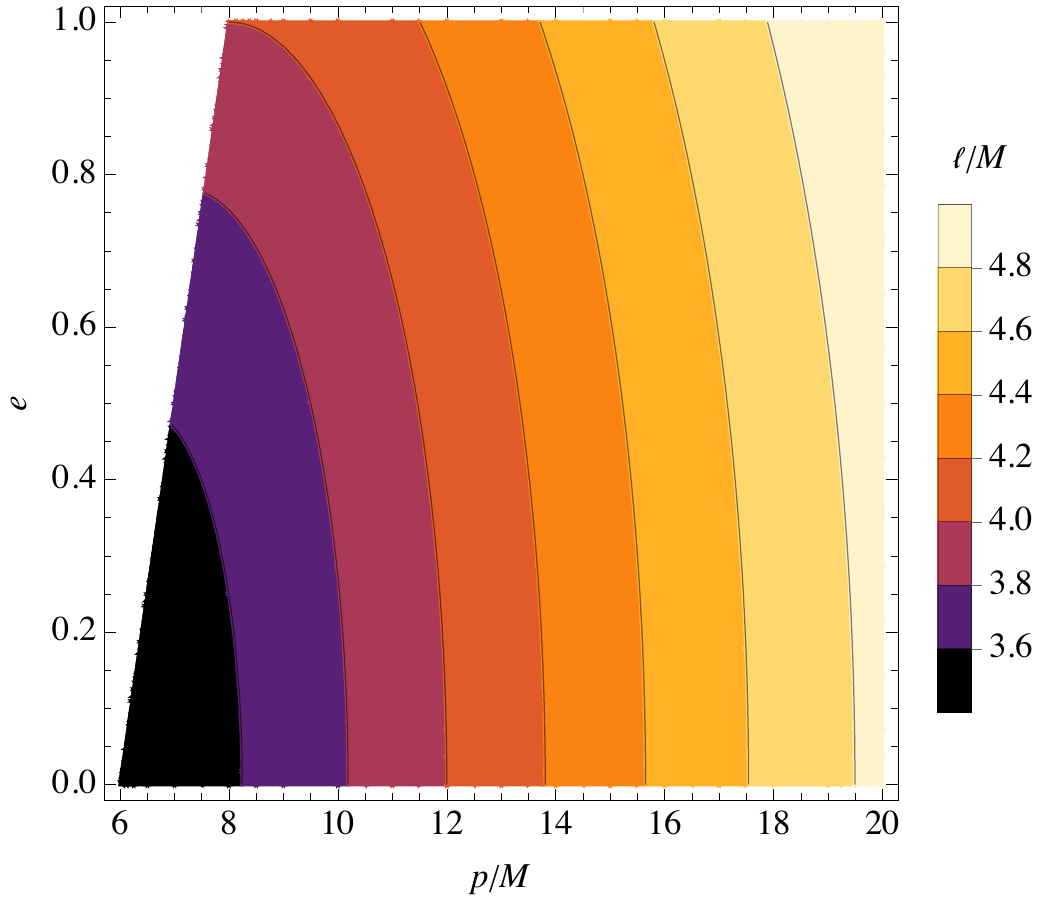}%
    \includegraphics[width=0.31\textwidth]{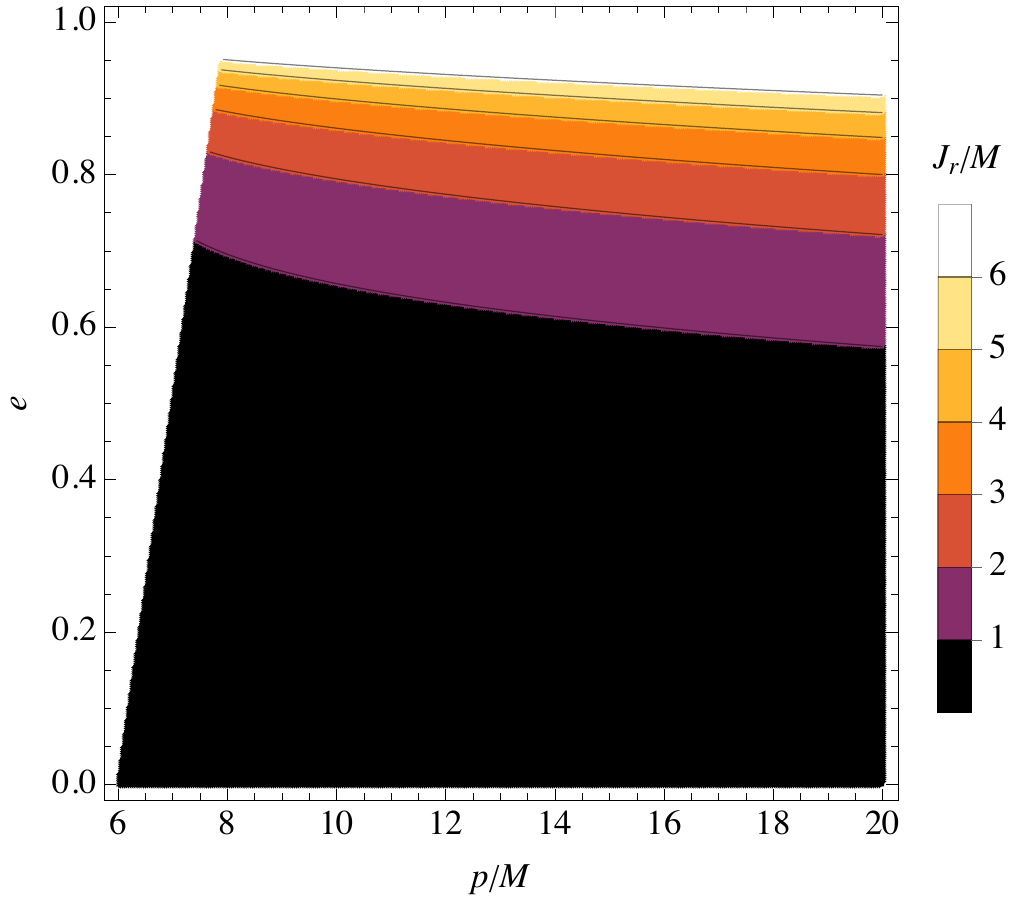}%
    \includegraphics[width=0.325\textwidth]{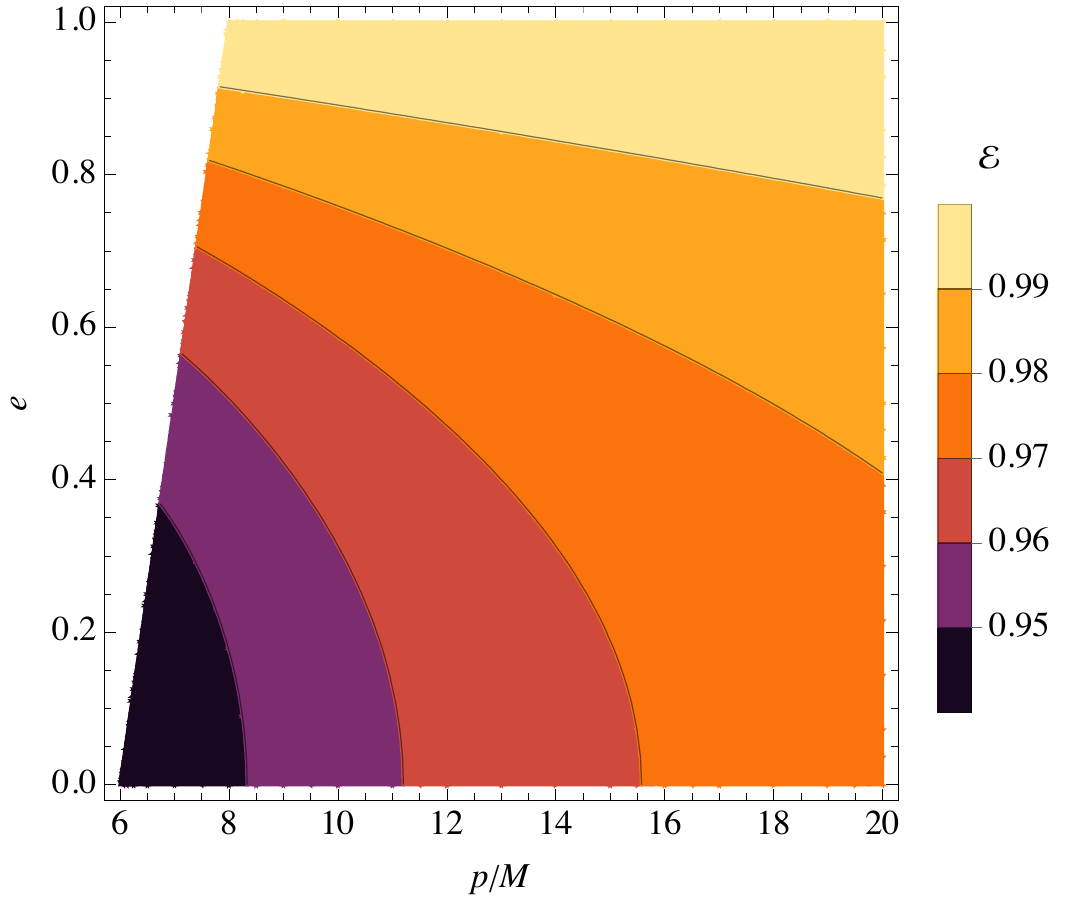}
    \caption{Exact $\ell, J_r, \cE$ as functions  of $e,p$ in respective order. The plot is cut off at the separatrix $p=(6+2e)M$. Note that $J_r$ diverges and $\cE$ goes to 1 in the $e\to 1$ limit.}
    \label{fig:epmap}
\end{figure*}

\begin{figure*}
    \centering
    \includegraphics[width=0.32\textwidth]{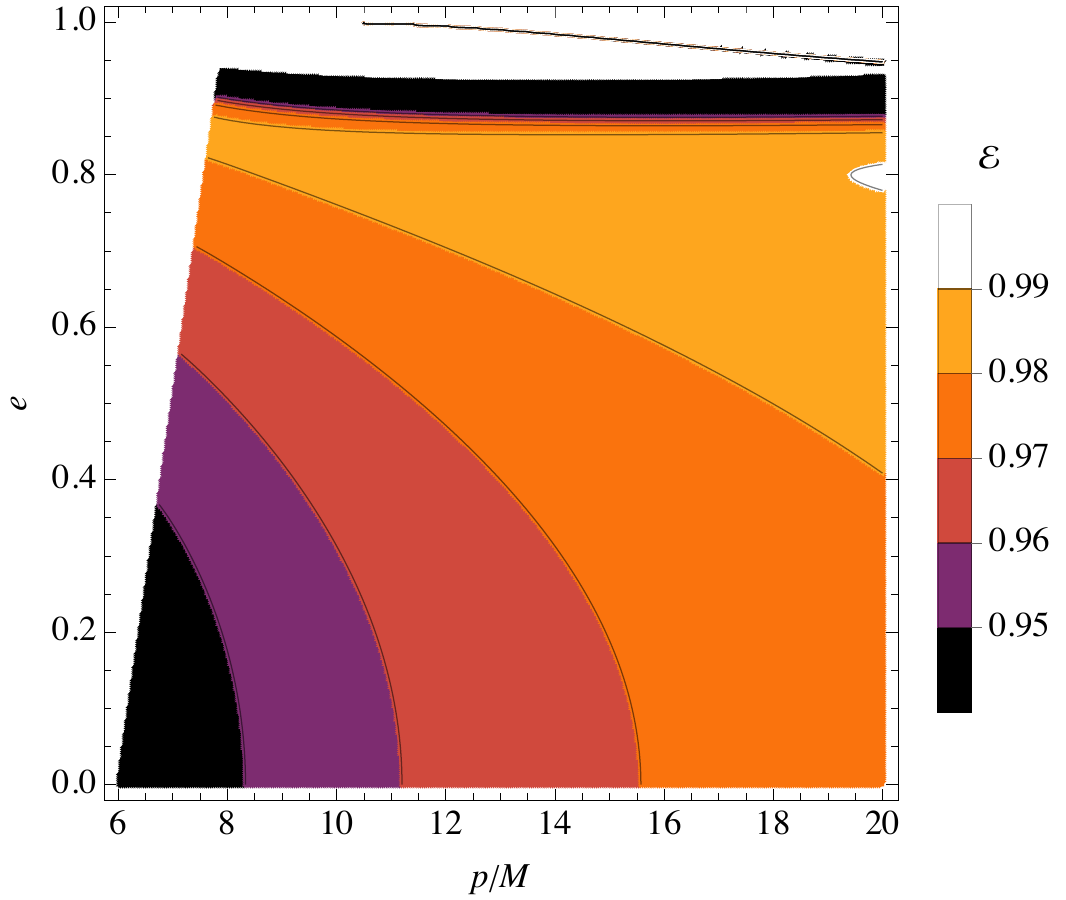}%
    \includegraphics[width=0.31\textwidth]{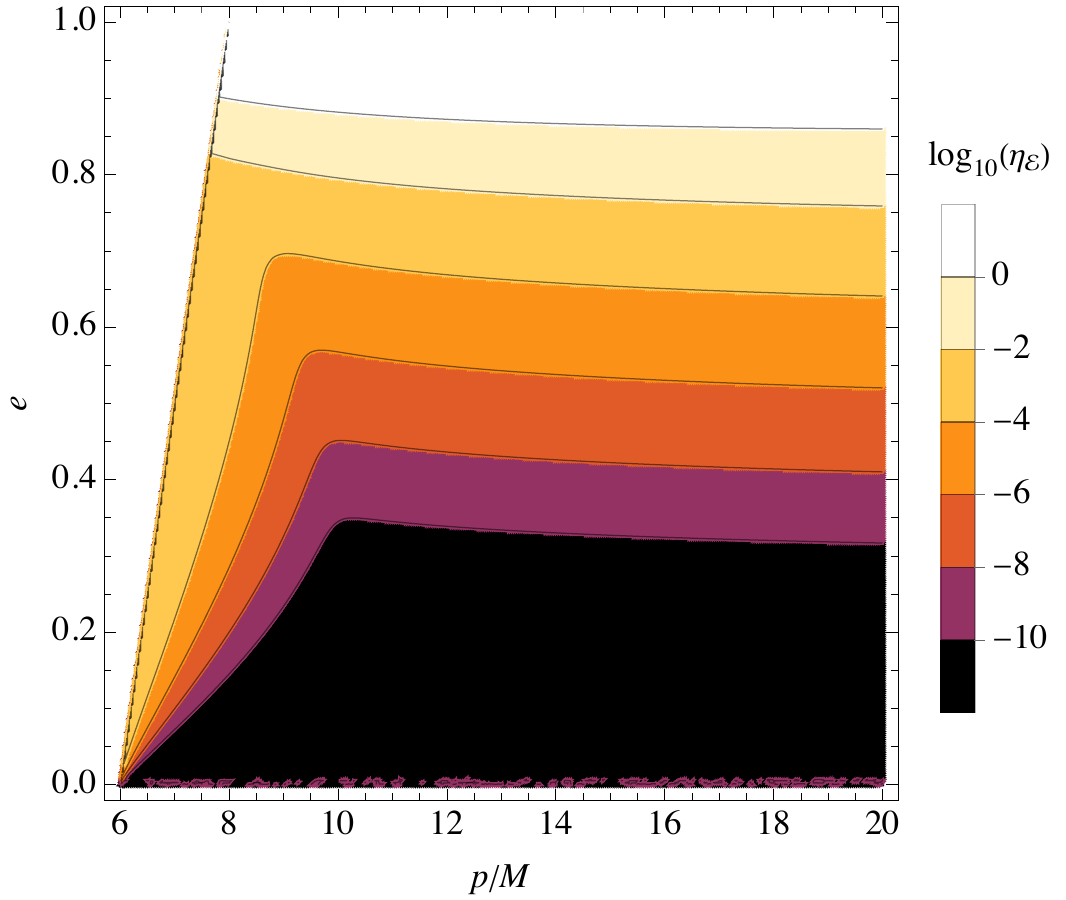}%
    \includegraphics[width=0.325\textwidth]{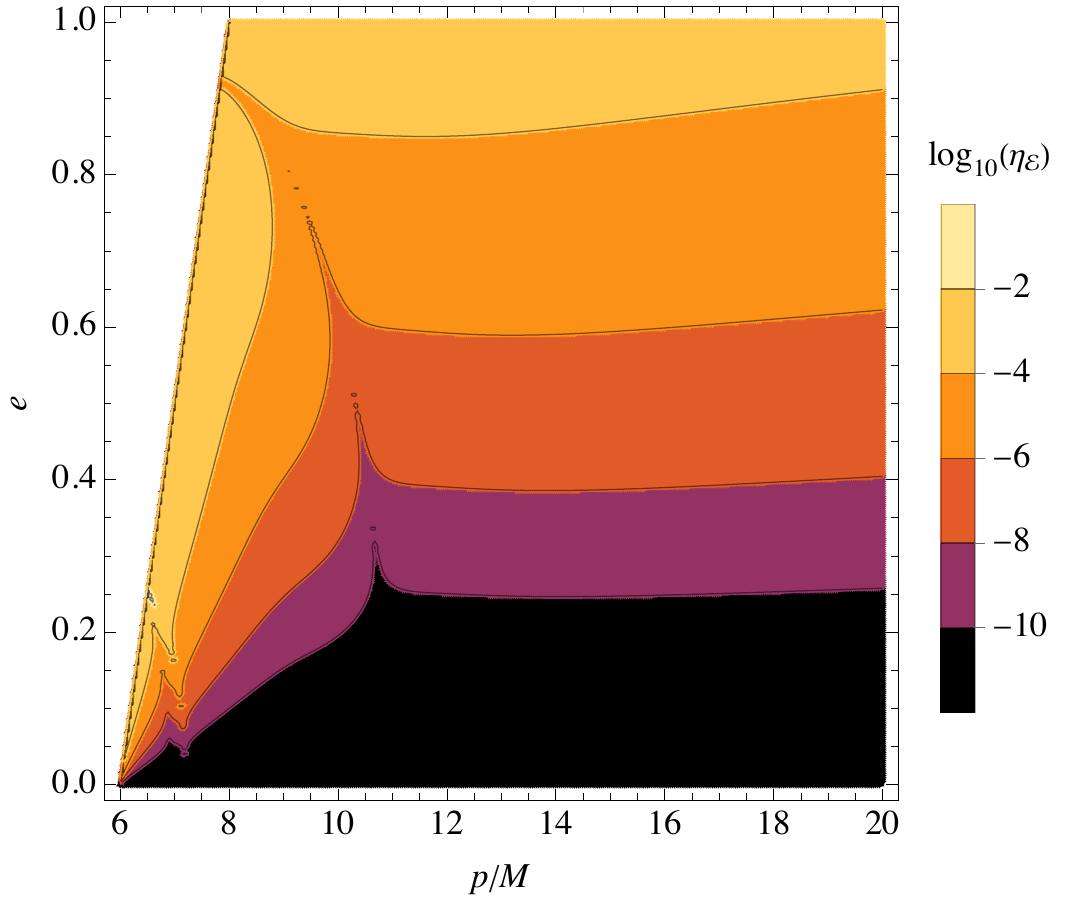}
    \caption{Energy $\cE(J_r,\ell)$ expanded to $\mathcal{O}(J_r^{10})$ plotted in $e,p$ space (left), the relative error of the expanded $\cE(J_r,\ell)$ (center), and the error of the $\mathcal{O}(J_r^5)$ Pad{\'e} resummation discussed in the text (right). While the Taylor expansion obviously has a convergence radius at $e\gtrsim 0.8$, the Pad{\'e} resummation is still able to provide an excellent global approximation for the Hamiltonian.}
    \label{fig:errs}
\end{figure*}

If one cares about observations over many periods, then one should require mainly the accuracy of the turning points and of the Hamiltonian (the derivatives of which determine, for instance, the average rates of pericentre precession and so on). This is investigated in Figures \ref{fig:epmap} and \ref{fig:errs}. Figure \ref{fig:epmap} presents the ``essential map'' of the $e-p$ space with respect to $\cE,\ell,J_r$ obtained either from exact relations or high-precision numerical integration. Figure \ref{fig:errs} then presents a comparison of the expanded formulas for the Hamiltonian with the exact picture. To do so, I define the relative error of the ``non-circular part'' of the Hamiltonian as
\begin{align}
    \eta_{\cE} = \Big|\frac{\cE_{\rm expand}(J_r(e,p),\ell(e,p)) - \cE_{\rm exact}(e,p)}{\cE_{\rm exact}(e,p) - \cE_{\rm exact}(e=0,p)} \Big|\,.
\end{align}

As is apparent from Figure \ref{fig:errs}, the expanded form of the Hamiltonian becomes divergent around $e\sim 0.8$. To improve the convergence, I devised a Pad{\'e} approximant as follows. I took the six terms that appear in the Hamiltonian by taking the expansion up to $n=10$ or $\mathcal{O}(J_r^5)$. Then I required that this expansion agrees with the Taylor expansion of the function
\begin{align}
    \cE_{\rm Pad} = \frac{\cE_{\rm c} + P_1 J_r +P_2 J_r^2+P_3 J_r^3}{1+ P_4 J_r +P_5 J_r^2+P_3 J_r^3}\,, \label{eq:Epad}
\end{align}
where the highest-order terms $P_3 J_r^3$ in the nominator and denominator are the same because I require that $\cE \to 1$ as $J_r \to \infty$. This can be easily seen to be true by considering that $J_r \to \infty$ as $e\to 1$ (see also Appendix \ref{app:asymp}). The coefficients $P_1,P_2,P_3,P_4,P_5$ are then uniquely determined by matching their $\mathcal{O}(J_r^5)$ expansion to the analytical expansion of $\cE$.The resulting approximant provides a Hamiltonian that has sub-percent level errors globally and can be understood simply as ``the'' action-angle Hamiltonian for bound orbits in most practical applications. 

Other asymptotics at the separatrix and at $e\to 1, p>8$ that could be used to refine this approximant are discussed in Appendix \ref{app:asymp}.

\subsection{Accuracy of quadrupole flux}

It is not entirely easy to measure the convergence of the orbital-shape harmonics in a meaningful way, since the orbital shapes are coordinate-dependent expressions with little direct physical meaning. However, I devised the following mock example that aims to demonstrate the usefulness of the formulas for gravitational-wave astrophysics. I assume that our coordinate system is aligned with the orbital plane of the motion and compute the Newtonian quadrupole of the orbiting particle
\begin{align}
    & Q_{xx} = \mu r^2 \ssq \ph\,,\\
    & Q_{xy} = \mu r^2 \sin \ph \cos \ph \,\\
    & Q_{yy} = \mu r^2 \csq \ph \,,
\end{align}
where $\mu$ is the small mass of the particle orbiting on the geodesic, and all the other $Q_{ij}$ components are zero. The lowest-order post-Newtonian formula for the power radiated in gravitational waves due to \citet{Peters:1963ux} then reduces to
\begin{align}
    P_{Q} = \left\langle \frac{2}{15} \left(\dddot{Q}_{xx}^2+3\dddot{Q}_{xy}^2+\dddot{Q}_{yy}^2\right) \right\rangle \,.
\end{align}
The expressions for the quadrupole components can be fully reconstructed from angle coordinates by using the expansions in eq. \eqref{eq:vpsir} and \eqref{eq:phpsiph}, and the property $\dot{\psi}^a = \Omega^a, a=r,\ph,\th$. The time-average can then be obtained for non-resonant orbits by the property
\begin{align}
\begin{split}
    &\lim_{T\to \infty}\frac{1}{T} \int_0^T\!\! f(\psi^r(t),\psi^\ph(t),\psi^\th(t)) \d t \\
    &= \frac{1}{(2 \pi)^3} \int_{[0,2\pi]^3}\!  f(\psi^r,\psi^\ph,\psi^\th) \d^3 \psi\,. \label{eq:avg}
\end{split}
\end{align}
Finally, the ``exact'' referential $P_Q$ can in principle be obtained by numerically integrating the geodesic and the corresponding average over the orbit. However, a naive numerical implementation of this strategy actually provides a \textit{worse} result than the analytical expansions in much of the phase space because of the oscillatory character of the integrals involved. It is instead beneficial to use the following trick. I first eliminate all the time derivatives in the formula for $P_Q$ by applying the equations of motion and express the result only in terms of $v,\ph$ and orbital elements. Then I use the fact that one can change integration variables in equation \eqref{eq:avg} to obtain \citep[see, e.g.][]{Drasco:2003ky,Witzany:2019nml}
\begin{align}
    & \lim_{T\to \infty}\frac{1}{T} \int_0^T\!\! f(v(t),\ph(t)) \d t 
    \\& = \frac{1}{2 \pi T_r} \int_{[0,2\pi]^2}\!  \frac{f(v,\ph)}{\dot{v}(v)} \d v \d \ph\,,
    \\ & T_r \equiv \int_0^{2\pi} \frac{\d v}{\dot{v}(v)}\,,
\end{align}
where when $f=P_Q$ the final expression can be evaluated numerically.

\begin{figure*}
    \centering
    \includegraphics[width=0.32\textwidth]{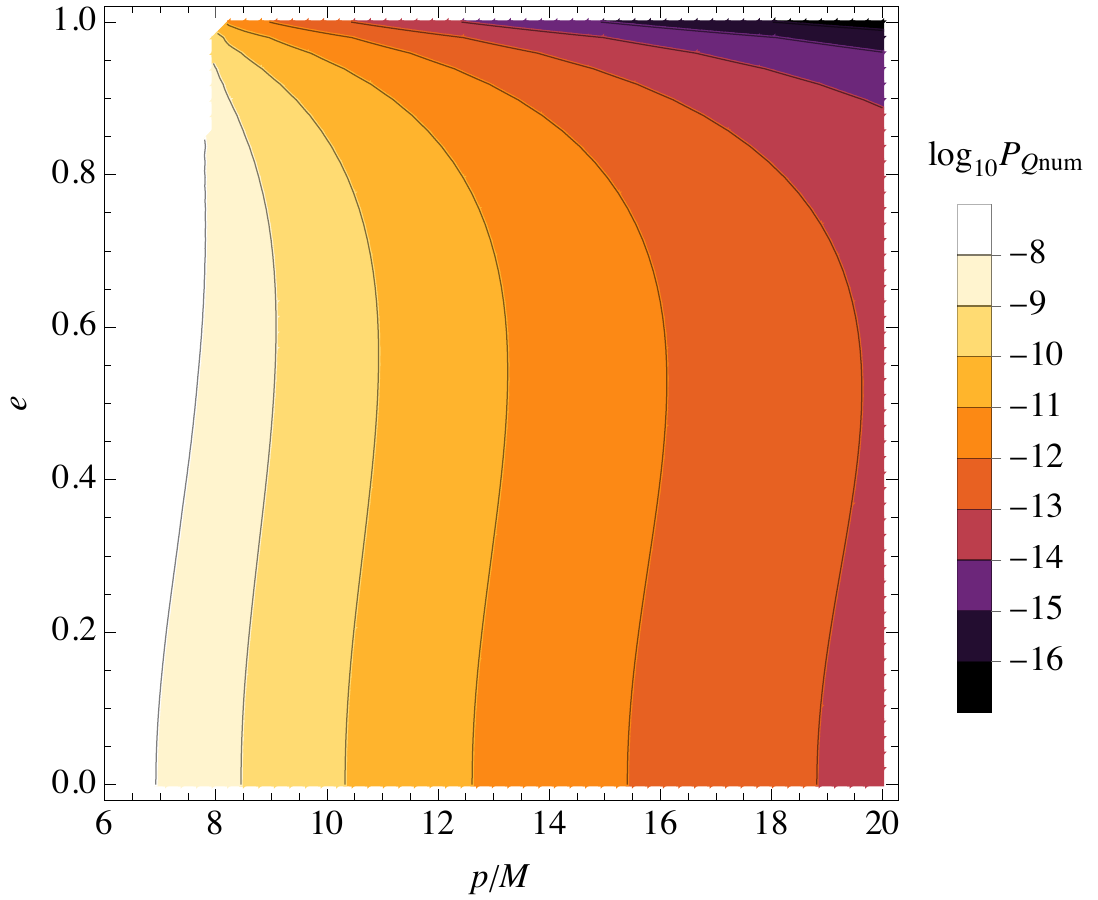}%
    \includegraphics[width=0.31\textwidth]{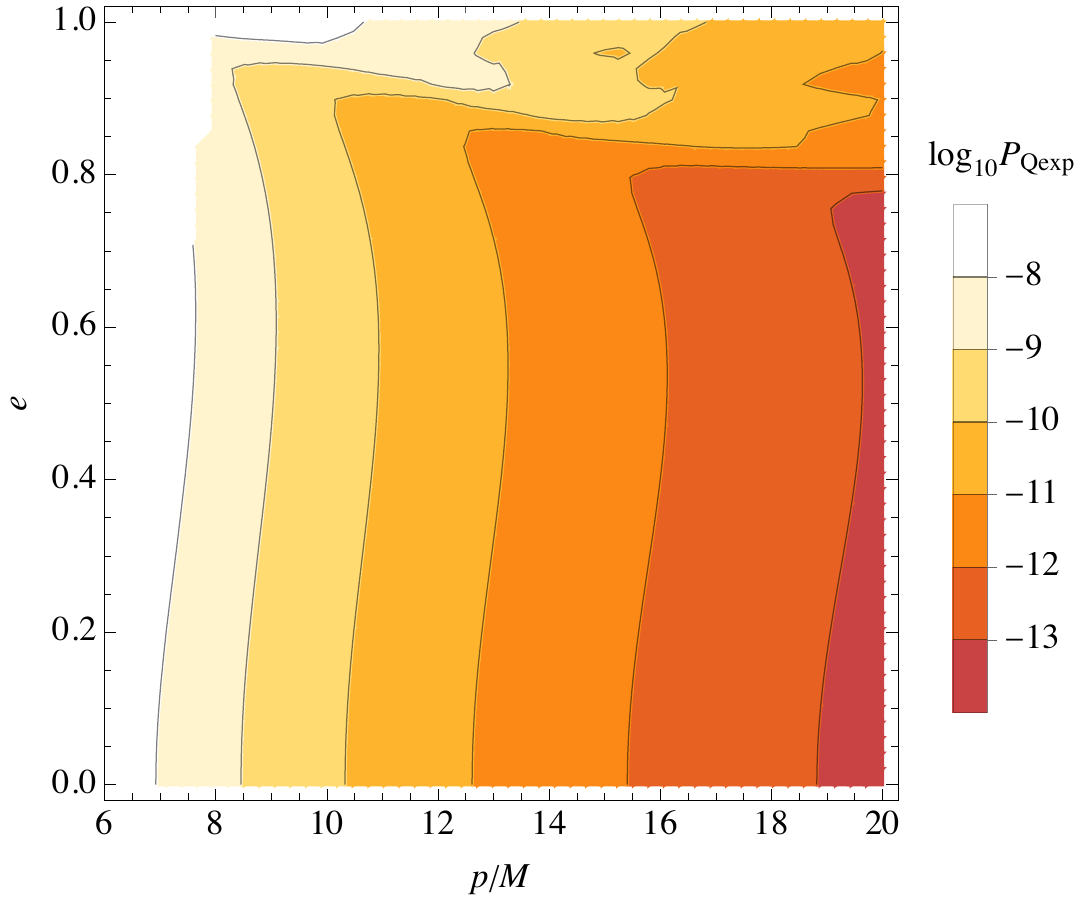}%
    \includegraphics[width=0.325\textwidth]{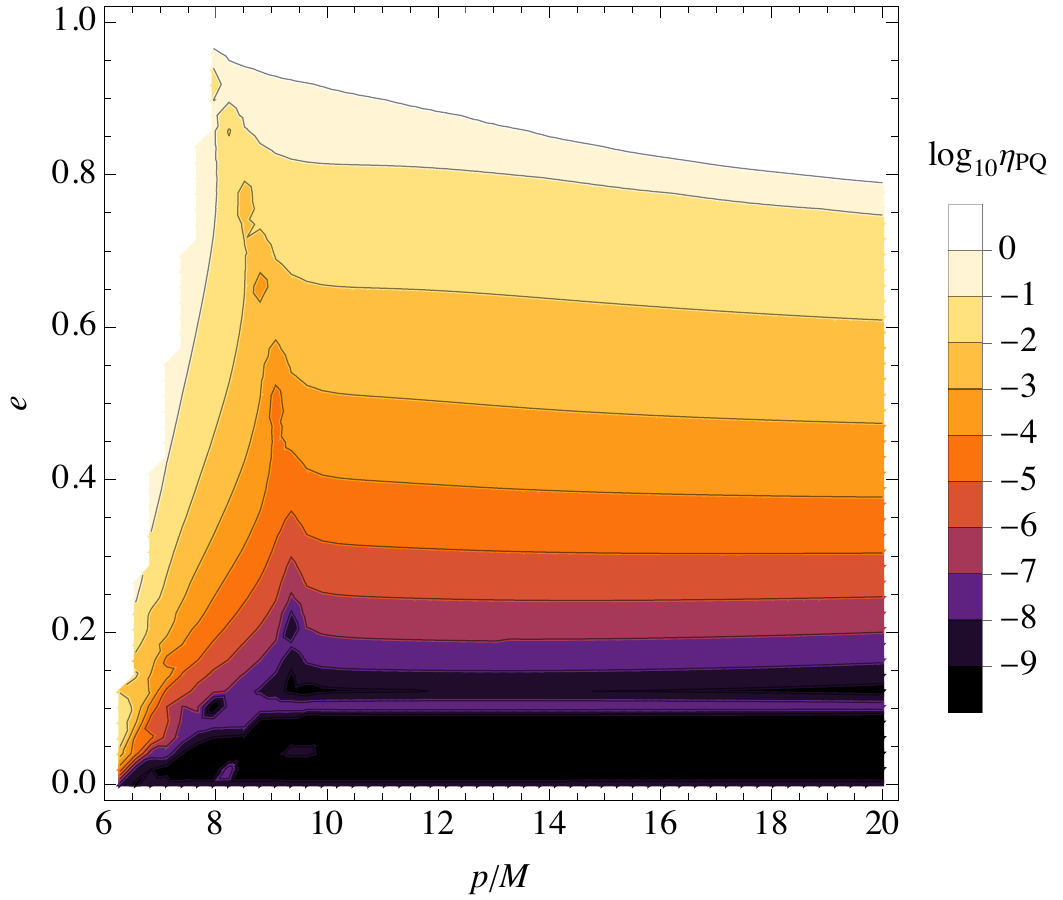}
    \caption{Quadrupole radiation power $P_{Q}$ computed numerically (left) and by the AA expansion (center). The relative difference between the two $\eta_{PQ}$ is then given on the right. Similarly to the other examples, the expansion seems to diverge near $e\sim 0.8$.}
    \label{fig:PQs}
\end{figure*}

Finally, the $P_{Q\rm exp}$ quadrupole flux predicted by the $n=8$ expanded action-angle solution can be compared with the exact (numerical) flux $P_{Q\rm num}$ as plotted in Fig. \ref{fig:PQs}. The story is quite similar to the expansion of the Hamiltonian; the expansion fares very well for small eccentricities but seems to have a limit of convergence at $e\sim 0.8$. In principle, the convergence could be again improved by observing that $P_Q\to 0$ as $e\to 1,\, J_r \to \infty$ and by using an appropriate Pad{\'e} approximant. However, this is out of the scope of the current paper.

\section{Discussion and outlooks} \label{sec:discuss}

The herein presented formalism constructs AA coordinates and thus also general solutions to the equations of motion for moderately eccentric bound geodesics even very close to the limit of stability near a non-rotating Schwarzschild black hole. However, there are questions of efficiency, usefulness, context, and extensions of this work that need to be discussed before I conclude the paper.

{\em Do we need actions?} AA coordinates play an important role in Hamiltonian perturbation theory and related near-identity transforms of the phase space variables. However, in gravitational-wave inspirals the perturbation is not Hamiltonian. More precisely, the infinite number of degrees of freedom of the disturbed gravitational field are not practical to include into a Hamiltonian description of the inspiral; this then makes the resulting dynamics ``dissipative''. In that case, one does not need to parametrize the phase-space by the actions, the only thing that is needed for closed-form near-identity transforms are the time-homogeneous phases, the ``angles'' of the AA variables. 

This approach was fleshed out for the case of extreme mass ratio inspirals by \citet{VanDeMeent:2018cgn}, where the authors in fact used Darwin's eccentricity and semi-latus rectum $e,p$ instead of actions (see also the applications of this formalism by \citet{McCart:2021upc, Lynch:2021ogr}). When implementing such an inspiral formalism in a concrete coordinate system, the formulas will be typically much simpler when different orbital elements than actions are used and when formulas such as equations \eqref{eq:psir}-\eqref{eq:psiph} are instead employed to derive the angle variables. In fact, the harmonics $\sigma_{ij},\theta_{ij}$ needed for the transformation from angle coordinates and thus also for the near-identity transforms can be obtained by a (non-expanded) discrete Fourier transform of the integrands and by trivial integration, similar to the methods employed by \citet{Hopper:2015jxa}. From the properties of the coefficients, it is obvious that by a discrete Fourier transform from $L$ samples of the integrands, one is obtaining results that are valid at least to order $e^L, J_r^{L/2}$ while retaining much better convergence properties. This is a direction that I hope to explore in the future.
 
{\em Actions are gauge-invariant.} Nevertheless, it should be stressed that actions can be very useful for other purposes. The definition of actions is coordinate-invariant and time-parametrization invariant, subject only to lattice transforms such as the one to Delaunay-type variables given in Section \ref{subsec:delaunay}. As a result, one can easily combine the results of various formalisms in action-angle coordinates without any need of further coordinate transforms. For example, I was able to directly check my results for consistency against the post-Newtonian expansions of the geodesic Hamiltonian as given by \citet{Bini:2020wpo}. Furthermore, one can also add the mass-ratio-proportional terms from Table IX from \citet{Bini:2020wpo} to the herein presented geodesic Hamiltonian to obtain an approximate description of conservative binary dynamics at finite mass ratio. Finally, the $J_a(t)$ curves corresponding to inspirals are gauge-invariant and amount to an extremely useful tool for comparison between different approaches to the relativistic two-body problem.

{\em Is the expansion worth it?} It is good to note that while the herein derived expansions are in principle analytical, the terms quickly become very large and summing them may become less efficient than a direct numerical integration and/or numerical inversion. The coefficients in the coordinate-independent $\cE-J_r-\ell$ relations swell slower with growing $n$ than the ones appearing in coordinate-dependent relations such as $v(\psi^r)$ or $\th (\psi^\th,\psi^r)$. For example, the $n=8$ post-circular $\langle P_{Q} \rangle$ formula can be slower to evaluate than $\langle P_{Q} \rangle$ obtained by numerical integration unless an extensive ``offline'' step is taken to express the $\psi^a$ averages analytically. However, as already noted in the Introduction, the matched-filtering method to be used by LISA data analysis will require the evaluation of an extremely large number of inspiral waveforms. As a result, even a modest speedup is worth extensive implementation and analytical work in this case. 

{\em Correspondence with \citet{Polcar:2022quad}.} This paper was written while I also contributed to the paper of \citet{Polcar:2022quad}. In that paper we also construct AA coordinates for geodesics in Schwarzschild space-time in a post-circular expansion, but under a different parametrization and using a different formalism, the so-called Lie series. The geodesics are then perturbed by a faraway axisymmetric gravitating source, and this perturbation is then absorbed by a near-identity transform so that we can compute inspirals of the test particles into the central black hole. The parametrization employed can be understood as the Schwarzschild limit of the so-called Carter-Mino time \citep[see also Appendix \ref{app:param}]{Carter:1968rr,Mino:2003yg}. As such, the angle coordinates in \citet{Polcar:2022quad} are different from those derived in this paper. Also, compared to the direct expansion of the solution of the Hamilton-Jacobi equation employed here, the Lie-series method becomes slightly more complex at higher order. On the other hand, the Lie series method can be used even if the solution of the Hamilton-Jacobi equation is unknown. As such, the papers provide a complementary view on AA coordinates and their properties in Schwarzschild space-time.

{\em What about Kerr?} The Schwarzschild case can be understood as a ``warmup'' for the truly generic astrophysical case: the motion in the field of a rotating isolated black hole also known as the Kerr space-time. Kerr geodesics and their AA coordinates will be the concern of a second paper paper in this series (currently in preparation). Unlike in the case of Schwarzschild space-time, the Kerr space-time is only axially symmetric and the AA Hamiltonian is a function of $\ell_{\rm(z)}, \ell, J_r$ instead of just $\ell, J_r$. There is a similar orbital parametrization to the Darwin elements $p,e$, with an additional parameter being the orbital inclination with respect to the spin axis $\iota$ \citep{Schmidt:2002qk}. The post-circular expansion can then be carried out with respect to circular equatorial orbits with respect to powers of $e,\iota$. However, the relation between energy and the orbital elements is less straightforward than in the Schwarzschild case, lower order of expansion can be reached, and the formalism is generally more complicated. It was thus highly important to first explore the expansion and its intricacies in the Schwarzschild case as presented in this paper.

\section*{Acknowledgements}
I would like to thank Maarten van de Meent, Lukáš Polcar, and Philip Lynch for providing feedback on drafts of this paper. This work was supported by European Union’s Horizon 2020 research and innovation programme under grant agreement No 894881.

\section*{Data Availability}

All the computations needed in this paper are implemented in two Mathematica notebooks, which generate several output files. All of these are available at \texttt{github.com/VojtechW/Action-Angle-Schwarzschild} and as Supplemental files to this paper.



\bibliographystyle{mn2e}
\bibliography{Lit} 




\appendix

\section{Other time parameters} \label{app:param}

 \subsection{Proper time}
 
 The Hamiltonian generating proper time evolution of geodesics in static, spherically symmetric space-times reads
 \begin{align}
     H_{\tau} = \frac{1}{2} g^{\mu\nu}p_\mu p_\nu =  \frac{1}{2} \left( -\frac{p_t^2}{a} + \frac{p_r^2}{b} + \frac{p_\th^2}{c} + \frac{p_\ph^2}{c \ssq \th} \right)\,,
 \end{align}
 where compared to the $t$-parametrization, the $t$-coordinate is now just a phase-space coordinate with the canonically conjugate momentum $p_t = -\cE$. 
 I denote the constant value of the Hamiltonian as $H_{\tau} = h$, with the physical on-shell value being $h = -1/2$ (however, $h$ now plays the role of a separation constant that cannot be set to $-1/2$ identically). The Hamilton-Jacobi equation corresponding to this Hamiltonian now has an almost identical solution with $S = h (\tau - \tau_0) - \cE(t - t_0) + \ell_{\rm (z)}(\ph - \ph_0)) + \tilde{S}_{(R)} + S_{(\th)}$. Now $S_{(\th)}$ is the identical one as given in equation \eqref{eq:sth} and
 \begin{align}
    & \tilde{S}_{(R)} = \int \! \sqrt{\frac{b}{a}\left[\cE^2 - b \left(\frac{\ell^2}{c} -2 h \right)\right]} \d R\,,
 \end{align}
 that is $\tilde{S}_{(R)}|_{h = -1/2} = S_{(R)}$. The definition of $J_\th,J_\ph$ is independent of the parametrization and we can add the action $J_t = p_t = -\cE$ (note that the definition is different for $J_t$ since the motion is unbound in $t$.) The definition of $J_R$ is then modified only by using the $h$-dependent $\tilde{S}_{(R)}$ instead of ${S}_{(R)}$. The angle variables $\tilde{\psi}$ corresponding to proper-time parametrization are then obtained as 
 \begin{align}
     & \tilde{\psi}^r = \frac{1}{\partial J_R/\partial h}\frac{\partial \tilde{S}_{(R)}}{\partial h}\,,\\
     & \tilde{\psi}^\th = \arctan\left(\frac{\sqrt{\ell^2 \ssq \th - \ell_{(z)}^2}}{\ell \cos\th}\right) -\frac{\partial J_R/\partial \ell}{\partial J_R/\partial h} \frac{\partial \tilde{S}_{(R)}}{\partial h} \,,  \\
    & \tilde{\psi}^\ph = \ph + \arctan\left( \frac{\ell_{(z)} \cos \th}{\sqrt{\ell^2 \ssq \th - \ell_{(z)}^2}} \right) + \mathrm{sign}(\ell_{(z)}) \tilde{\psi}^\th, \\
    & \tilde{\psi}^t = t - \frac{\partial \tilde{S}_{(R)}}{\partial \cE} + \frac{\partial J_R}{\partial \cE} \tilde{\psi}^R\,.
 \end{align}

 \subsection{Carter-Mino time}

Notice that one can take the proper-time Hamiltonian and change the parametrization into a parameter $\lambda$ by
\begin{align}
    H_{\lambda} = f \left(H_\tau + \frac{1}{2} \right)\,,
\end{align}
where $f$ is any non-zero phase-space function and the parameter $\lambda$ then fulfils $\d \lambda/\d \tau = 1/f$. In the case of spherically symmetric space-times it is beneficial to choose $f = c(R)$, since then the Hamiltonian becomes fully separable in $R$ and $\th$. The value of the Hamiltonian is a constant of motion denoted as $H_\lambda = \Lambda$, and its on-shell value is 0 (again, this is kept as a separation constant that is not zero identically). This leads to the solution of the Hamilton-Jacobi equation of the form $S = h (\tau - \tau_0) - \cE(t - t_0) + \ell_{\rm (z)}(\ph - \ph_0)) + \bar{S}_{(R)} + S_{(\th)}$ with
 \begin{align}
    & \bar{S}_{(R)} = \int \! \sqrt{\frac{b}{a}\left[\cE^2 - b \left(\frac{\ell^2 - \Lambda}{c} +1 \right)\right]} \d R\,,
 \end{align}
The other steps are then analogous as in the proper-time parametrization and the dependence on $h$. The formulas for the angle coordinates $\bar{\psi}$ corresponding to the $\lambda$ parametrization then also have the analogous form to the formulas for $\tilde{\psi}$ above
\begin{align}
     & \bar{\psi}^r = \frac{1}{\partial J_R/\partial \Lambda}\frac{\partial \bar{S}_{(R)}}{\partial \Lambda}\,,\\
     & \bar{\psi}^\th = \arctan\left(\frac{\sqrt{\ell^2 \ssq \th - \ell_{(z)}^2}}{\ell \cos\th}\right) -\frac{\partial J_R/\partial \ell}{\partial J_R/\partial \Lambda} \frac{\partial \bar{S}_{(R)}}{\partial \Lambda} \,,  \\
    & \bar{\psi}^\ph = \ph + \arctan\left( \frac{\ell_{(z)} \cos \th}{\sqrt{\ell^2 \ssq \th - \ell_{(z)}^2}} \right) + \mathrm{sign}(\ell_{(z)}) \bar{\psi}^\th, \\
    & \bar{\psi}^t = t - \frac{\partial \bar{S}_{(R)}}{\partial \cE} + \frac{\partial J_R}{\partial \cE} \bar{\psi}^R\,.
 \end{align}
The name ``Carter-Mino'' parametrization can be clarified by comparing with the angle variables in an analogous parametrization in Kerr space-time \citep{Schmidt:2002qk,Fujita:2009bp, vandeMeent:2019cam}.

\section{Post-circular expansion in spherically symmetric space-times} \label{app:AAspher}

 \subsection{Hamiltonian in terms of actions}
 
 I start by determining the turning points of motion (roots in the integrand in $S_{(R)}$) by expanding
 \begin{align}
    & \cE = \cE_{\rm c}(\ell) + \sum_{i=1}^{[n/2]} \kappa^{2i} \delta_i \cE  \,,  \; R_{1,2} = R_{\rm c}(\ell) + \sum_{k=1}^n \kappa^k \delta_k R_{1,2} \label{eq:ERexp}\,,
 \end{align}
 where $\delta_i \cE, \delta_k R_{1,2}$ are unknown corrections, $\kappa$ is a bookkeeping parameter$\sim \sqrt{J_R}\sim e$, and $n$ is a chosen expansion order (in the main text $\kappa$ is set to one for simplicity of expressions). The meaning of $\delta_i \cE$ is the energy change caused by some power of the radial action, later I will set $\delta_i \cE = \varepsilon_i(\ell)J_R^i$, but I keep the $\delta \cE$ notation for now. 
 
 Now the turning points are determined by solving
 \begin{align}
     & \cE^2 - B(R_{1,2};\ell) =0 \,,
 \end{align}
with the expansions \eqref{eq:ERexp}, plugged in. The radial fluctuations $\delta_k R_{1,2}$ can then be solved for in terms of $\delta_i \cE$ order by order (I choose a convention such that $R_1<R_2$) to obtain
\begin{align}
    &\delta_1 R_{1,2} = \mp 2 \sqrt{\frac{ \cE_{\rm c} \, \delta_1 \cE}{B^{''}}}\,, \; \delta_2 R_{1,2} = - \frac{2 \cE_{\rm c} \, \delta_1 \cE B^{(3)}}{3 B^{''2}}\,, \\
    & \delta_3 R_{1,2} = \mp \,... \nonumber\,,
\end{align}
where primes denote $R$ derivatives and $F^{(n)}\equiv \d^n F/\d R^n$. Additionally, all functions and derivatives are evaluated at $R=R_{\rm c}(\ell)$. The explicit expansions of $\delta_k R_{1,2}$ to order $k=10$ are given in a Mathematica notebook in the supplemental material. It generally holds that the odd contributions have signs alternating for the individual $R_1, R_2$ turning points and, on the other hand, the even expansions terms add the same ``shift'' contribution to $R_{1},R_2$. 

Now I take the expression for $J_R$ in eq. \eqref{eq:jr} and perform a substitution to an integration variable $\rho$ such that $R = (R_2-R_1)\rho + R_1$ to obtain the expansion
\begin{align}
    & \kappa^2 J_R = \frac{(R_2 - R_1)}{\pi} \int_0^1 \sum_{i=1}^n \sqrt{\rho(\rho-1)} \, \kappa^i \delta_i I(\rho) \d \rho\,, \label{eq:jrexp}\\
    & \delta_1 I = 2\sqrt{2 \cE_{\rm c} \,\delta_1 \cE A}\,,\\
    & \delta_2 I = \frac{2\sqrt{2}\, \cE_{\rm c} \,\delta_1 \cE(2 \rho - 1)\left(3A^{'} B^{''} + A B^{(3)}\right)}{3 \sqrt{A B^{''3}}} \,,\\
    & \delta_3 I = ... \nonumber\,,
\end{align}
where again all functions are evaluted at $R=R_{\rm c}(\ell)$. The $\kappa^2$ factor on the left-hand side of equation \eqref{eq:jrexp} appears necessarily due to the following behaviour of the right-hand side. Each of the terms of the expansion of the integral on the right-hand side can be computed exactly, and only odd terms $i=1,3,5,...$ terms end up as non-zero. Similarly, the expansion of the prefactor $R_1-R_2$ only has odd powers of $\kappa$ appearing in it, so the total product on the right-hand side only has even powers of $\kappa$ starting from $\kappa^2$. One is then left with an expansion for the action of the form
\begin{align}
     & J_R = \sum_{i = 0}^{[n/2]-1}  \kappa^{2i} k_i(\cE,R_{\rm c})\,,\label{eq:jrfinexp}\\
     & k_1 = \sqrt{\frac{\cE_{\rm c} \,\delta_1 \cE A}{8}}\,,\; k_2  = \,\frac{1}{\mathcal{C}}\left[36 \cE_{\rm c} A^2 B^{''3} \delta_2 \cE_{\rm c} +  \mathcal{D} \,\delta_1 \cE^2  \right]\,,\\
    & \mathcal{C} \equiv 144 B^{''3} \sqrt{2 \cE_{\rm c} \, \delta_1\cE \, A^3}   \,,\\
    \begin{split}
     & \mathcal{D} \equiv 18 A^2 B^{''3} + \cE_{\rm c}^2\Big[18 A A^{''} B^{'' 2} -9 A^{'2} B^{''2} 
     \\ & \phantom{\mathcal{D} \equiv} - 18 A A^{'} B^{''} B^{(3)} - 5A^2 (B^{(3)})^2 + 3 A^2 B^{''} B^{(4)}\Big] \,,
    \end{split} \\
     & k_3 = \, ... \nonumber\,.
\end{align}
Now this relation can be inverted perturbatively by setting $\delta_i \cE = \varepsilon_i(\ell)J_R^i$ and solving for the coefficients $\varepsilon_i(\ell) = \varepsilon_i( R_{\rm c}(\ell))$ order by order.

When the dust settles, one has the expression for the energy as
\begin{align}
    & \cE = \cE_{\rm c}(\ell) + \sum_{i=1}^{[n/2]}\kappa^{2i}J_R^i \eps_i(\ell)\,, \label{eq:Efin}\\
    & \eps_1 = \sqrt{\frac{B^{''}}{2 \cE_{\rm c}^2 A}}\,,\; \eps_2 = \frac{\cE_{\rm c}^2 \mathcal{F}  - 12A^2 B^{''3}}{48 \cE_{\rm c}^3 A^3 B^{''2}}\,,\\
    \begin{split}
     & \mathcal{F} \equiv 3A^{'2} B^{''2} + 6A\left(A^{'} B^{''} B^{(3)} -A^{''}B^{''2}\right) \\  & \phantom{\mathcal{F} \equiv} + A^2 \left(3B^{''} B^{(4)} - 5(B^{(3)})^2\right) \,,
    \end{split} \\
    & \eps_3 = ... \nonumber\,,
\end{align}
where expressions up to $\varepsilon_5$ (corresponding to $\sim\kappa^{10}$) can be again found in the supplemental material. Note that the expressions are purposefully parametrized by $R_{\rm c}, \cE_{\rm c}$ while assuming that the expression for $R_{\rm c}(\ell), \cE_{\rm c}(\ell)$ will be substituted into the final result. Now one can see that upon the substitution $\ell = J_\th + |J_\ph|$ into \eqref{eq:Efin}  we obtain a perturbative expression for the Hamiltonian $H_t$ in terms of AA coordinates, $H_t = \cE(J_R,J_\th,J_\ph)$.

For completeness, we can also perturbatively invert \eqref{eq:Efin} to obtain $J_R(\cE)$ as
\begin{align}
    & J_R = \sum_{i=1}^{[n/2]} \kappa^{2i} \left[\cE -\cE_{\rm c}(\ell)\right]^i j_{R i}(\ell)\,,  \label{eq:JE}\\
    & j_{R1} = \frac{1}{\eps_1} = \cE_{\rm c} \sqrt{\frac{2 A}{B^{''}}}\,,\; j_{R2} = - \frac{\eps_2}{\eps_1^3} = \frac{12 A^2 B^{''3} - \cE_{\rm  c}^2 \mathcal{F}}{12 \sqrt{A^3 B^{''7}}} \,,\\
    & j_{R3} = ... \nonumber \,.
\end{align}

\subsection{Angle variables} \label{appsub:angles}
I now construct the angle part of the variables using the indirect method \eqref{eq:indir} in the post-circular expansion. The only reason to use the indirect formula here is that for various technical reasons it is simpler to make a derivative of the integral $S_{(R)}$ with respect to $\cE$ rather than $J_R$. I start by expanding the integrals
\begin{align}
    & I_{(\cE)} \equiv \frac{\partial S_{(R)}}{\partial \cE} = \int  \cE \sqrt{\frac{A}{\cE^2 - B}} \d R\,, \label{eq:IrE} \\
    &I_{(\ell )}\equiv \frac{\partial S_{(R)}}{\partial \ell} = -\int\!\frac{\ell C}{\sqrt{\cE^2 - B}} \d R \label{eq:Irell}\,,\\
    &C(R) \equiv \frac{b }{c \sqrt{A}} = \frac{\sqrt{ab}}{c} \,.
\end{align}
I set $\psi^R$ to zero when at the smallest value of the radial coordinate (the pericenter), that is, at $R=R_1$. This corresponds to the integrals \eqref{eq:IrE} and \eqref{eq:Irell} computed from a lower bound at $R_1$. Additionally, I transform to an auxiliary phase variable $\xi$ with the properties
\begin{align}
    & R = \frac{1}{2}(R_2 - R_1)\cos \xi + \frac{1}{2}(R_2 + R_1) \,,\\
    & \xi = \arccos \frac{2R - (R_2 + R_1)}{R_2 - R_1} = 2 \arcsin \sqrt{\frac{R-R_1}{R_2 - R_1}}\,.
\end{align}
Or, in terms of the integrable variable used for the action, $\rho = \ssq(\xi/2)$. The use of $\xi$ makes the expanded integrands only trigonometric polynomials with no square roots and fractions, which accelerates symbolic processing. Specifically, it is advantageous to store the expansion in terms of coefficients of a trigonometric Fourier transform over $\xi \in (0,2\pi)$ and use parallelization to carry out (exact) integration and algebraic simplifications coefficient by coefficient. 

Using the expansions of turning points and energy as in equations \eqref{eq:ERexp} and \eqref{eq:Efin} one then obtains
\begin{align}
    & I_{(\cE)} =  \left[\left(\sum_{i = 0}^{[n/2]-1} \!\!\!\!\! \kappa^{2i} \nu_{(\cE) i} J_R^{i}\right) \xi + \sum_{k=1}^{n-1} \sum_{j=1}^{n-1} \kappa^k J_R^{k/2} \iota_{(\cE)kj} \sin(j \xi)    \right] ,\\
    & I_{(\ell)} =  \left[\left(\sum_{i = 0}^{[n/2]-1} \!\!\!\!\! \kappa^{2i} \nu_{(\ell) i} J_R^{i}\right) \xi + \sum_{k=1}^{n-1} \sum_{j=1}^{n-1} \kappa^k J_R^{k/2} \iota_{(\ell)kj} \sin(j \xi)    \right], \\
    & \nu_{(\cE)0} = \cE_{\rm c} \sqrt{\frac{2 A}{B^{''}}}\,,\; \nu_{(\ell)0} = \ell C \sqrt{\frac{2}{B^{''}}}\,,\\
    & \nu_{(\cE,\ell)1} = ... \,,\\
    & \iota_{(\cE,\ell)kj} \neq 0 \; \mathrm{iff} \; j\leq k \; \& \; k+j \; \mathrm{even}\,, \nonumber\\
    & \iota_{(\cE)11} = \frac{2^{1/4} \cE_{\rm c}  (A B^{(3)} - 3 A^{'} B^{''})}{ 3 (A^3 B^{''7})^{1/4}}\,,\\
    & \iota_{(\ell)11} = \frac{2^{1/4} \ell (C B^{(3)} - 6 C^{'} B^{''})}{3 (A B^{''7})^{1/4}}\,,\\
    & \iota_{(\cE,\ell)22} =  ...\,. \nonumber
\end{align}
Another way to obtain $\sum \kappa^{2i} \nu_{(\cE)i}$ (and an important consistency check) is to take a derivative of eq. \eqref{eq:JE} with respect to $\cE$ and substituting eq. \eqref{eq:Efin}. 

Now, using eq. \eqref{eq:psir} it is easy to find 
\begin{align}
    & \psi^R = \xi + \sum_{i=1}^{n-1} \sum_{j=1}^{n-1} \kappa^i J_R^{i/2} \Xi_{ij} \sin(j\xi) \,, \label{eq:psiRexp}\\
    & \Xi_{ij} \neq 0 \; \mathrm{iff} \; j\leq i \; \& \; i+j \; \mathrm{even}\,, \nonumber\\
    & \Xi_{11} = \frac{ (A B^{(3)} - 3 A^{'} B^{''})}{2^{1/4} 3 (A B^{''})^{5/4}}\,, \; \Xi_{22} = ... \nonumber \,.
\end{align}
 The fact that only $\sin(k \xi)$ terms only appear in the expansion follows from reversibility and choosing $\xi=\psi^R = 0$ at the turning point. The inversion of this equation to $\xi(\psi^R)$ can then be computed perturbatively as
\begin{align}
    & \xi = \psi^R + \sum_{i=1}^{n-1} \sum_{j=1}^{n-1} \kappa^i J_R^{i/2} \Psi_{ij} \sin(j\psi^R) \,, \label{eq:xiexp}\\
    & \Psi_{ij} \neq 0 \; \mathrm{iff} \; j\leq i \; \& \; i+j \; \mathrm{even}\,, \nonumber\\
    & \Psi_{11} = -\frac{ (A B^{(3)} - 3 A^{'} B^{''})}{2^{1/4} 3 (A B^{''})^{5/4}}\,, \; \Psi_{22} = ...  \,.
\end{align}
The vanishing/non-vanishing properties of $\Xi_{ij},\,\Psi_{ij}$ is integral to being able to keep the expansion manageable to high order in $J_R$.

Let us now turn our attention to the angle variables $\psi^\th, \psi^\ph$. This is largely analogous to the case in Newtonian gravity as given, e.g., in \citet{binney2011galactic}. Using eqs \eqref{eq:psith} and \eqref{eq:psiph} we have 
\begin{align}
    & \psi^\th = \arctan\left(\frac{\sqrt{\ell^2 \ssq \th - \ell_{(z)}^2}}{\ell \cos\th}\right) + \chi\,,\\
    & \psi^\ph = \ph + \arctan\left( \frac{\ell_{(z)} \cos \th}{\sqrt{\ell^2 \ssq \th - \ell_{(z)}^2}} \right) + \mathrm{sign}(\ell_{(z)}) \chi\,, \label{eq:apsiphexp}
\end{align}    
where, again, different branches of $\arctan$ have to be used in each of the two expression to obtain functions smooth around $\th=\pi/2$. Specifically, the expression for $\psi^\ph$ requires $\arctan(0^+) = \arctan(0^-)$, and the expression for $\psi^\th$ should have $\arctan(+\infty) = \arctan(-\infty)$. The auxiliary function $\chi$ is then given as
\begin{align}
    \begin{split}
        & \chi(\psi^R,J_R,\ell) \equiv I_{(\ell)}\Big|_{\xi=\xi(\psi^R)} - \frac{\partial J_R}{\partial \ell} \psi^R   \\
        & \phantom{\chi} = \sum_{k=1}^{n-1} \sum_{j=1}^{n-1} \kappa^k J_R^{k/2} X_{kj} \sin(j \psi^R) ,
    \end{split} \\
    & X_{ij} \neq 0 \; \mathrm{iff} \; j\leq i \; \& \; i+j \; \mathrm{even}\,, \nonumber\\
    & X_{11} = \frac{2^{1/4} \ell \left[C A^{'} (A B^{'' 11})^{1/4} - 2A^{5/4} C^{'}B^{'' 11/4}\right]}{A^{3/2} B^{'' 7/2} } \,, \\ 
    & X_{22} = ...\,, \nonumber
\end{align}
where $\partial J_R/\partial \ell$ cancels all the non-oscillating terms in $I_{(\ell)}$. Consequently, $\chi$ expresses the purely oscillatory influence of $\psi^r$ on the $\th,\ph$ coordinates. Note that the integration constants were chosen so that
\begin{align}
    \th = \pi/2,\, \ph=0,\, \xi=0 \,(R=R_1) \leftrightarrow \psi^R = \psi^\th = \psi^\ph = 0\,.
\end{align}

It is now easy to invert for $\th$ as
\begin{align}
    \cos \th = \sqrt{1 - \frac{\ell_{\rm (z)}^2}{\ell^2}} \, \cos(\psi^\th - \chi)\,, \label{eq:thpsithsp}
\end{align}
where a sign ambiguity during the inversion is fixed by the sign choice already made in eq. \eqref{eq:jth}.

The last thing to express is the dependence of the azimuthal angle $\ph$ on the action-angle variables. Equations \eqref{eq:psiphexp} and \eqref{eq:thpsithsp} yield
\begin{align}
    & \ph = \psi^\ph  - {\rm sign}(\ell_{\rm(z)}) \chi - \arctan \left[\frac{\ell_{\rm(z)}}{\ell} \cot(\psi^\th - \chi) \right]\,. \label{eq:phps}
\end{align}
That is, the only difficult part of evaluating the relation between $\th,\ph$ and the action-angle variables is the function $\chi$. In addition to these, one then needs to also evaluate the fundamental frequencies as detailed in Section \ref{subsec:funfrek}

This finishes this section of the Appendix, since all that one needs is to specify the metric functions $a,b,c$ (or $A,B,C$) and $R_{\rm c}(\ell), \cE_{\rm c}(\ell)$ to explicitly obtain the expanded action-angle coordinates in any spherically symmetric space-time.


\section{Asymptotics of radial action and energy} \label{app:asymp}

\subsection{High $J_r$, $e\to 1$}

The coefficients of the $J_r$-expansion become increasingly complex with higher orders with possibly decreasing yields in convergence. Hence, it is advantageous to also interpolate the Hamiltonian with the $J_r \to \infty$ limit in order to obtain a good qualitative approximation everywhere in parameter space. \citet{Bini:2020wpo} obtained the $J_r(\mathcal{E};\ell), \mathcal{E}(J_r;\ell)$ relation in a $\ell\to \infty$ expansion to order $1/\ell^{12}$. I take the additional limit $J_r \to \infty$ of their $\cE(J_r;\ell)$ relation to obtain the expansion (cf. Table XI in \cite{Bini:2020wpo})
\begin{align}
\begin{split}
    & \cE =  1 - \frac{M^2}{2 J_r^2} 
    \\ & + \frac{\ell^3}{J_r^3}\left[ \frac{M^2}{\ell^2} -\frac{3 M^4}{\ell^4} -\frac{35M^6}{4 \ell^6} - \frac{231 M^8}{4 \ell^8} - \frac{32175 M^{10}}{64 \ell^{10}} +... \right] 
    \\ & + \frac{\ell^4}{J_r^4} \left[\frac{-3M^2}{2\ell^2} + \frac{87 M^4}{8 \ell^4} + \frac{51 M^6}{4 \ell^6} + \frac{189 M^8}{2 \ell^8} + \frac{55911 M^{10}}{64 \ell^{10}} + ...\right]
    \\ & + ...\,, \label{eq:ElargeJ}
\end{split}
\end{align}
where each $\sim 1/J_r^i$ apart from $i=2$ has a nontrivial ongoing expansion in $1/\ell$ to at least $1/\ell^{12}$ order. I thus conjecture that the $\sim 1/J_r^2$ term is exact in $\ell$ everywhere where the $1/\ell$ expansion converges (which is likely only true up to some $\ell \gtrsim 4M$). Obviously, the large $J_r$ limit coincides with the $\cE \to 1$ limit for the $J_r(\cE)$ relation, for which I obtain (cf. equations (13.20) and (13.22) in \cite{Bini:2020wpo})
\begin{align}
    J_r = \frac{M}{\sqrt{2(1 - \cE)}} + \ell \left(-1 + \frac{3M^2}{\ell^2} + \frac{35 M^4}{4 \ell^4}+... \right) + \mathcal{O}(\sqrt{1 - \cE}).
\end{align}
Once again, I conjecture that the leading-order term is exact to all orders in the $1/\ell$ expansion everywhere where the expansion is convergent. 

Let us also examine the asymptotic of energy given by equation \eqref{eq:Epe} in the $e\to 1$ limit. This yields
\begin{align}
    \cE \to 1 + \frac{2 M^2 (e - 1)}{\ell (\ell + \sqrt{\ell^2 -16 M^2})} + \mathcal{O}\left((e-1)^2\right)\,.
\end{align}
Considering equation \eqref{eq:ElargeJ} this means that, \textit{assuming $\ell>4M$}, the $J_r \to \infty$ and $e\to1$ limits coincide in orbital space with 
\begin{align}
    (1 - e) \to  \frac{\ell (\ell + \sqrt{\ell^2 -16 M^2})}{4 J_r^2 M} + \mathcal{O}\left((e-1)^2\right)\,.
\end{align}
Once again, these can be used to interpolate the post-circular formulas to the large-eccentricity asymptotics.

\subsection{Near-separatrix}

Another point to consider are the asymptotics around the separatrix. For this one notices that the integral \eqref{eq:srep} can be evaluated in closed form at $p=(6+2e)M$ to yield
\begin{align}
\begin{split}
   & \Js \equiv J_r|_{p=(6+2e)M} = 
\frac{M}{\pi  \sqrt{e^4-10 e^2+9}} \times
    \\& \Bigg[ 2 \left(e^2+7\right) \arctan \left(\sqrt{\frac{2 e}{1-e}}\right)
   -2 \sqrt{2e(1-e)} (e+3)
   \\
   & -8\sqrt{2(1-e^2)} {\rm arctanh} \left(\sqrt{\frac{e}{e+1}}\right)
   \Bigg]
    \end{split}
\end{align}
Along the separatrix we  can then express 
\begin{align}
    e|_{\rm sep.}(\ell) = \frac{\ell  \left(\ell +2 \sqrt{\ell ^2-12 M^2}\right)-12 M^2}{\ell ^2+4 M^2}
\end{align}
and thus we have the separatrix action parametrized by $\ell$
\begin{align}
\begin{split}
    & \Js(\ell) = 
    \\& \frac{1}{\pi  \sqrt{\lambda ^2 \left(384-\lambda  \left(\lambda
    \left(13 \lambda ^2+14 \lambda  l+188\right)+160 l\right)\right)+9216}} \times
    \\& \Bigg[ \left(\lambda ^2 (\lambda  (3 \lambda +l)+68)+448\right) \arctan\left(
   \sqrt{\frac{3 \lambda }{l-2 \lambda }}\right)
    \\& -2 \left(\lambda ^2+16\right) \sqrt{128-2
   \lambda ^2 (\lambda  (\lambda +l)+4)} {\rm arccoth}\left(
   \sqrt{\frac{2(\lambda +l)}{3\lambda }}\right)
   \\& -\left(\lambda ^{3/2} \left(l+2\lambda\right)+24\right) \sqrt{16
   l-\lambda  (\lambda  (2 \lambda +l)+16)}
    \Bigg]\,,
\end{split}
\\ & l \equiv \frac{\ell}{M} \,, \lambda \equiv \sqrt{l^2 - 12}\,.
\end{align}
Now it is also easy to obtain the energy along the separatrix
\begin{align}
    \cE|_{\rm sep.}(\ell) = \sqrt{\frac{l (36+l(l + \lambda))-12 \lambda}{54 l}}\,.
\end{align}
Note that the orbits on the separatrix are simply the homoclinic infinite-zoom-while orbits as discussed at the beginning of Section \ref{sec:schw} and this expression is then identical to $\cE_{\rm h}$ given in equation \eqref{eq:Eh}.
Finally, since the separatrix is defined by $\partial \cE/\partial J_r = \Omega^r = 0$, we obtain the near-separatrix form of the AA Hamiltonian
\begin{align}
    \cE(J_r,\ell) = \cE|_{\rm sep.}(\ell) + \mathcal{O}\left((J_r - \Js(\ell))^2\right)\,.
\end{align}

Naive attempts at incorporating this information into an approximant is non-trivial since one has to smoothly switch between the separatrix asymptotics for $\ell \in [\sqrt{12 M},4M]$ and then to $J_r \to \infty$ asymptotics for $\ell>4M$. None of my attempts to do so have managed to produce a more satisfying result than the simple Pad{\'e} approximant in eq. \eqref{eq:Epad}.


\bsp	
\label{lastpage}
\end{document}